\documentclass[aip,apl,10pt,numerical,superscriptaddress,reprint,twocolumn]{revtex4-1}

\usepackage{amsmath}
\usepackage{amsfonts}
\usepackage{amssymb}
\usepackage{graphicx}
\usepackage{color}
\usepackage{hyperref}
\hypersetup{colorlinks=true,allcolors=blue}

\newcommand{\KET}[1]{\vert #1\rangle}
\newcommand{\BRA}[1]{\langle #1\vert}
\newcommand{\OP}[2]{\KET{#1}\BRA{#2}}
\newcommand{\PRO}[1]{\OP{#1}{#1}}
\newcommand{\aH}{\hat{a}_{\text{H}}}
\newcommand{\aHd}{\aH^\dagger}
\newcommand{\aV}{\hat{a}_{\text{V}}}
\newcommand{\aVd}{\aV^\dagger}
\newcommand{\n}{\left\langle n\right\rangle}

\begin{document}


\title[Title option]{Time-dependent switching of the photon entanglement type using a driven quantum emitter-cavity system} 



\author{T. Seidelmann}
\email[Author to whom correspondence should be addressed: T. Seidelmann: ]{tim.seidelmann@uni-bayreuth.de}
\affiliation{Lehrstuhl f{\"u}r Theoretische Physik III, Universit{\"a}t Bayreuth, 95440 Bayreuth, Germany}
\author{D. E. Reiter}
\email{doris.reiter@uni-muenster.de}
\affiliation{Institut f{\"u}r Festk{\"o}rpertheorie, Universit{\"a}t M{\"u}nster, 48149 M{\"u}nster, Germany}
\author{M. Cosacchi}
\affiliation{Lehrstuhl f{\"u}r Theoretische Physik III, Universit{\"a}t Bayreuth, 95440 Bayreuth, Germany}
\author{M. Cygorek}
\affiliation{Heriot-Watt University, Edinburgh EH14 4AS, United Kingdom}
\author{A. Vagov}
\affiliation{Lehrstuhl f{\"u}r Theoretische Physik III, Universit{\"a}t Bayreuth, 95440 Bayreuth, Germany}
\affiliation{ITMO University, St. Petersburg, 197101, Russia}
\author{V. M. Axt}
\affiliation{Lehrstuhl f{\"u}r Theoretische Physik III, Universit{\"a}t Bayreuth, 95440 Bayreuth, Germany}

\date{\today}

\begin{abstract}
The cascaded decay in a four-level quantum emitter is a well established mechanism to generate polarization entangled photon pairs, the building blocks of many applications in quantum technologies. The four most prominent maximally entangled photon pair states are the Bell states. In a typical experiment based on an undriven emitter only one type of Bell state entanglement can be observed in a given polarization basis. Other types of Bell state entanglement in the same basis can be created by continuously driving the system by an external laser. In this work we propose a protocol for time-dependent entanglement switching in a four-level quantum emitter--cavity system that can be operated by changing the external driving strength. By selecting different two-photon resonances between the laser-dressed states, we can actively switch back and forth between the different types of Bell state entanglement in the same basis as well as between entangled and nonentangled photon pairs. This remarkable feature demonstrates the possibility to achieve a controlled, time-dependent manipulation of the entanglement type that could be used in many innovative applications.
\end{abstract}


\maketitle 



Entangled qubits are the building blocks for fascinating applications in many innovative research fields, like  quantum cryptography\cite{Gisin:02,Lo_quantum_cryptography}, quantum communication\cite{duan_quantum_comm,Huber_overview_2018}, or quantum information processing and computing\cite{pan:12,Bennett:00,Kuhn:16,Zeilinger_entangled}. Besides possible applications, the phenomenon of entanglement is also important from a fundamental point of view, being a genuine quantum effect. Especially attractive realizations of two entangled qubits are polarization entangled photon pairs, because they travel at the speed of light and are hardly influenced by the environment\cite{Orieux_entangled}.

The most prominent maximally entangled states, established for polarization entangled photons pairs, are the four Bell states
\begin{subequations}
\begin{equation}
\label{eq:def_pol_ent_state}
\KET{\Phi_\pm} = \frac{1}{\sqrt{2}}\left( \KET{HH} \pm \KET{VV} \right) ,
\end{equation}
\begin{equation}
\label{eq:def_bell_ent_state}
\KET{\Psi_\pm} = \frac{1}{\sqrt{2}}\left( \KET{HV} \pm \KET{VH} \right) ,
\end{equation}
\end{subequations}
where $H$ and $V$ denote horizontally and vertically polarized photons, respectively. The order corresponds to the order of photon detection: In a $\Phi$ Bell state ($\Phi$BS) the first and second detected photon exhibit the same polarization, whereas in a $\Psi$ Bell state ($\Psi$BS) the two detected photons have exactly the opposite polarization.

A well established mechanism for the creation of these maximally entangled Bell states is the cascaded decay that takes place in a four-level quantum emitter (FLE) after an initial excitation. Such a FLE can be realized by a variety of systems including F-centers, semiconductor quantum dots or atoms\cite{edamatsu2007entangled,Freedman_FLE_atom, Wen_theory_atomic_system, Park_FLE_atom}. Employing a FLE, $\Phi$BS entanglement in the chosen basis of linearly polarized photons was demonstrated for various conditions in both theoretical and experimental studies\cite{Seidelmann2019,Different-Concurrences:18, Phon_enhanced_entanglement,Jahnke2012, heinze17,BiexcCasc_Carmele,Stevenson2006,Young_2006,Muller_2009,Huber_PRL_2018, Wang_2019,Liu2019, Bounouar18,dousse:10,winik:2017, entangled-photon1,Fognini_2019, entangled-photon2,Hafenbrak, Biexc_FSS_electrical_control_Bennett, EdV,Troiani2006,stevenson:2012, Benson_2000_QD_cav_device}. In contrast, $\Psi$BS entanglement in the same linearly polarized basis has only been predicted in the case of continuous laser driving \cite{munoz15,Seidelmann_QUTE_2020}. For the driven FLE laser-dressed states emerge, which have been observed experimentally \cite{Ardelt_exp_const_driving,Hargart_exp_const_driving}. By embedding the FLE inside a microcavity with cavity modes tuned in resonance with the desired emission process,  certain two-photon emission processes between the laser-dressed states can be favored \cite{munoz15,Seidelmann_QUTE_2020}. The emerging type and degree of entanglement depends strongly on the dominant two-photon emission path between the laser-dressed states, which, in turn can be tuned by the external driving strength \cite{Seidelmann_QUTE_2020}.

Based on these findings, we propose a protocol for time-dependent entanglement switching using a driven FLE-cavity system. Simply changing the external driving strength in a step-like manner enables one to actively switch between the generation of $\Phi$BS and $\Psi$BS entanglement as well as between entangled and nonentangled photon pairs. Therefore, different entangled states can be generated from the same source without further processing the photons to change the entanglement, e.g., by wave plates.


\begin{figure}
\centering
\includegraphics[width=0.8\columnwidth]{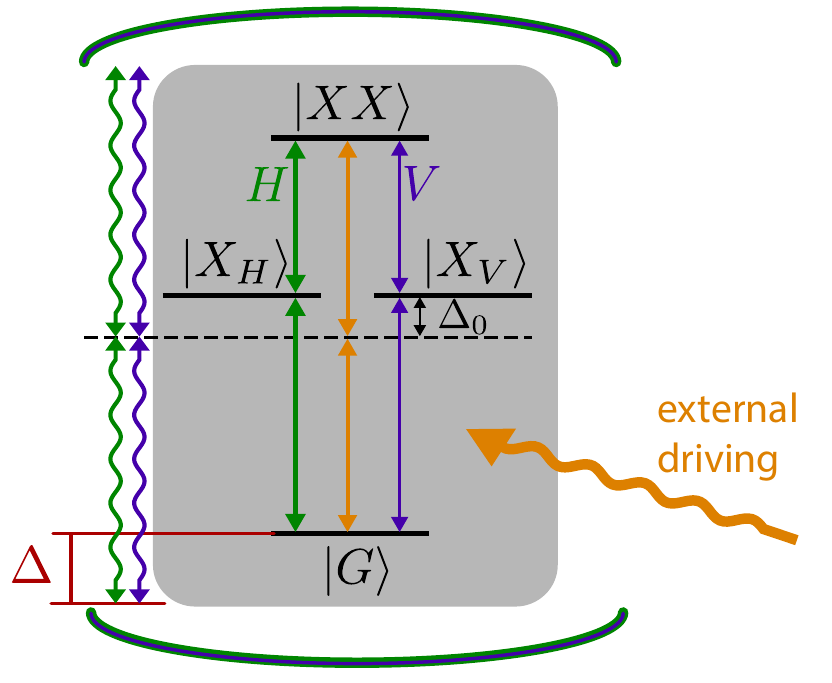}
\caption{
Sketch of the driven FLE-cavity system. The FLE consists of the states $\KET{G}$, $\KET{X_\text{H/V}}$, and  $\KET{XX}$, which are coupled via optical transitions by horizontally/vertically polarized light (green/purple straight arrows). The FLE is driven by an external laser at the two-photon resonance which results in a detuning of $\Delta_0$ to the intermediate states (orange arrows). The FLE is embedded into a cavity with two energetically degenerate, but orthogonal horizontally/vertically polarized cavity modes (green/purple wavy arrows) detuned by $\Delta$ to the laser energy. 
} 
\label{fig:model}
\end{figure}

We consider an externally driven FLE-cavity system, which has been presented in detail in Refs.~\onlinecite{munoz15,Seidelmann_QUTE_2020}.
Figure~\ref{fig:model} depicts a sketch of this system. A generic FLE comprises the ground state $\KET{G}$, two degenerate intermediate single-excited states $\KET{X_\text{H/V}}$, and the upper state $\KET{XX}$. Typically, $\KET{XX}$ is not found at twice the energy of the single-excited states, but is shifted by the value $E_\text{B}$, e.g., in quantum dots $E_B$ is referred to as the biexciton binding energy \cite{Orieux_entangled,Mermillod:16}. Transitions between the FLE states that involve the state $\KET{X_\text{H/V}}$ are coupled to horizontally/vertically polarized light. If the $\KET{XX}$ state has been prepared \cite{entangled-photon1,Reiter_2014,reindl2017,hanschke2018} cascaded photon emission takes place when the FLE relaxes to its ground state resulting in the typical $\Phi$BS.

An external laser with driving strength $\Omega$ is used to excite the FLE. The laser frequency is adjusted such that the two-photon transition between the ground state $\KET{G}$ and $\KET{XX}$ is driven resonantly, resulting in a fixed energetic detuning $\Delta_0=E_\text{B}/2$ between the single-excitation transitions and the laser (cf., Fig.~\ref{fig:model}). The laser polarization is chosen to be linear with equal components of the $H$ and $V$ polarization. The FLE is placed inside a microcavity and coupled to its two energetically degenerate linearly polarized modes, $H$ and $V$. The energetic placement of the cavity modes is described by the cavity laser detuning $\Delta$, i.e., the difference between the cavity mode and laser energy. In typical set-ups, the fabrication process determines $\Delta$ and it cannot be changed afterwards. Accordingly, we fix the cavity laser detuning to $\Delta=0.8\Delta_0$. The coupling strength $g$ between cavity and FLE is assumed to be equal for all FLE transitions.

Furthermore, important loss processes, i.e., radiative decay with rate $\gamma$ and cavity losses with rate $\kappa$, are included using Lindblad-type operators \cite{Seidelmann_QUTE_2020,Lindblad:1976}. The time evolution of the statistical operator of the system and two-time correlation functions are calculated by numerically solving the resulting Liouville-von Neumann equation\cite{multi-time}. The system parameters for the calculations are displayed in Table~\ref{tab:Fixed_Parameters} \cite{munoz15,Seidelmann_QUTE_2020}. Initially the system is in the FLE ground state $\KET{G}$ without any cavity photons. For the Hamiltonian and details on the calculations we refer to Ref.~\onlinecite{Seidelmann_QUTE_2020}.

\begin{table}
\centering
\caption{Fixed system parameters used in the calculations.}
\label{tab:Fixed_Parameters}
\begin{ruledtabular}
\begin{tabular}{l c c}
Parameter & & Value\\
\hline
Coupling strength & $g$ & $0.051$~meV \\
Detuning & $\Delta_0$ & $20g=1.02$~meV \\
Cavity laser detuning & $\Delta$ & $0.8\Delta_0=0.816$~meV \\
Cavity loss rate & $\kappa$ & $0.1g/\hbar \approx 7.8$~$\mathrm{ns^{-1}}$\\
Radiative decay rate & $\gamma$ & $0.01g/\hbar \approx 0.78$~$\mathrm{ns^{-1}}$\\
\end{tabular}
\end{ruledtabular}
\end{table}

The entanglement characterization relies on the standard two-time correlation functions
\begin{equation}
\label{eq:G2}
G_{jk,lm}^{(2)}(t,\tau^\prime) = \left\langle \hat{a}_j^\dagger(t)\hat{a}_k^\dagger(t+\tau^\prime)\hat{a}_m(t+\tau^\prime)\hat{a}_l(t) \right\rangle
\end{equation}
with $\lbrace j,k,l,m\rbrace\in\lbrace H,V\rbrace$\cite{Different-Concurrences:18}. Here, $t$ is the real time of the first photon detection and $\tau^\prime$ the delay time between this detection event and the detection of the second photon. The operator $\hat{a}_\text{H/V}^\dagger$ creates one horizontally/vertically polarized cavity photon\cite{Note1}.
\footnotetext[1]{Note that in typical experiments the measurements are performed on photons which have already left the cavity. Nevertheless, when the out-coupling of light out of the cavity is considered to be a Markovian process, Eq.~\eqref{eq:G2} can be used to describe $G^{(2)}(t,\tau)$ as measured outside of the cavity [cf. Refs.~\onlinecite{Kuhn:16,Different-Concurrences:18}].}
In realistic two-time coincidence experiments the data is always obtained by averaging the signal over finite real time and delay time intervals. Consequently, we use averaged correlation functions that depend on the starting time of the coincidence measurement $t_0$, the used real time measurement interval $\Delta t$, and the delay time window $\tau$ (see also  Ref.~\onlinecite{Seidelmann_QUTE_2020}.).  

A measure to classify the entanglement is the two-photon density matrix $\rho^\text{2p}$, from which the resulting type of entanglement can be extracted directly from its form. In standard experiments $\rho^\text{2p}$ is reconstructed employing quantum state tomography\cite{QuantumStateTomography} and, consequently, it is obtained from the averaged correlation functions as detailed in Ref.~\onlinecite{Seidelmann_QUTE_2020}.

To quantify the degree of entanglement we use the concurrence $C$, which can be calculated directly from the two-photon density matrix \cite{Wootters1998,EdV,QuantumStateTomography,Seidelmann_QUTE_2020,Note2}.
\footnotetext[2]{ $C=\max\left\lbrace \sqrt{\lambda_1}-\sqrt{\lambda_2}-\sqrt{\lambda_3}-\sqrt{\lambda_4},0\right\rbrace$ where $\lambda_j\geq\lambda_{j+1}$ are the eigenvalues of $\rho^\text{2p}T{\rho^\text{2p}}^\ast T$ in decreasing order and $T$ is the antidiagonal matrix with elements $\left\lbrace -1,1,1,-1\right\rbrace$.}
Note that both, the two-photon density matrix and the concurrence, depend on the parameters of the coincidence measurements: $t_0$, $\Delta t$, and $\tau$. Throughout this article a delay time window $\tau=50$ ps is assumed\cite{StevensonPRL2008}.


Before presenting the switching protocol, we study the behavior of the constantly driven FLE-cavity system as a function of the driving strength for a fixed selected cavity laser detuning. The resulting type of entanglement and its degree depend on the cavity laser detuning $\Delta$ and the driving strength $\Omega$, as demonstrated in Ref.~\onlinecite{Seidelmann_QUTE_2020}. In particular, a high degree of $\Phi$BS or $\Psi$BS entanglement is only possible, when the cavity modes are close to or in resonance with a direct two-photon transition between the laser-dressed states of the FLE. In the present set-up we have fixed all frequencies and detunings, such that the only free tuning parameter is the driving strength $\Omega$. 

The constant driving of the FLE results in a mixing of the bare states $\KET{G}$, $\KET{X_\text{H/V}}$, and  $\KET{XX}$, such that the new eigenstates are the laser-dressed states, which we label by $\KET{U}$, $\KET{M}$, $\KET{N}$, and $\KET{L}$. Their respective energies are given by\cite{Seidelmann_QUTE_2020}
\begin{subequations}
\begin{eqnarray}
E_\text{U} &=&  \frac{1}{2} \left( \Delta_0 +\sqrt{\Delta_0^2+ 8\Omega^2} \right)\\
E_\text{M} &=& \Delta_0 \\
E_\text{N} &=& 0 \\
E_\text{L} &=&  \frac{1}{2} \left( \Delta_0 - \sqrt{\Delta_0^2+ 8\Omega^2} \right).
\end{eqnarray}
\end{subequations}
Both the state mixing and the energies depend on the driving strength $\Omega$, which we will now use to tune certain two-photon transitions in resonance with the cavity modes.

\begin{figure}
\centering
\includegraphics[width=\columnwidth]{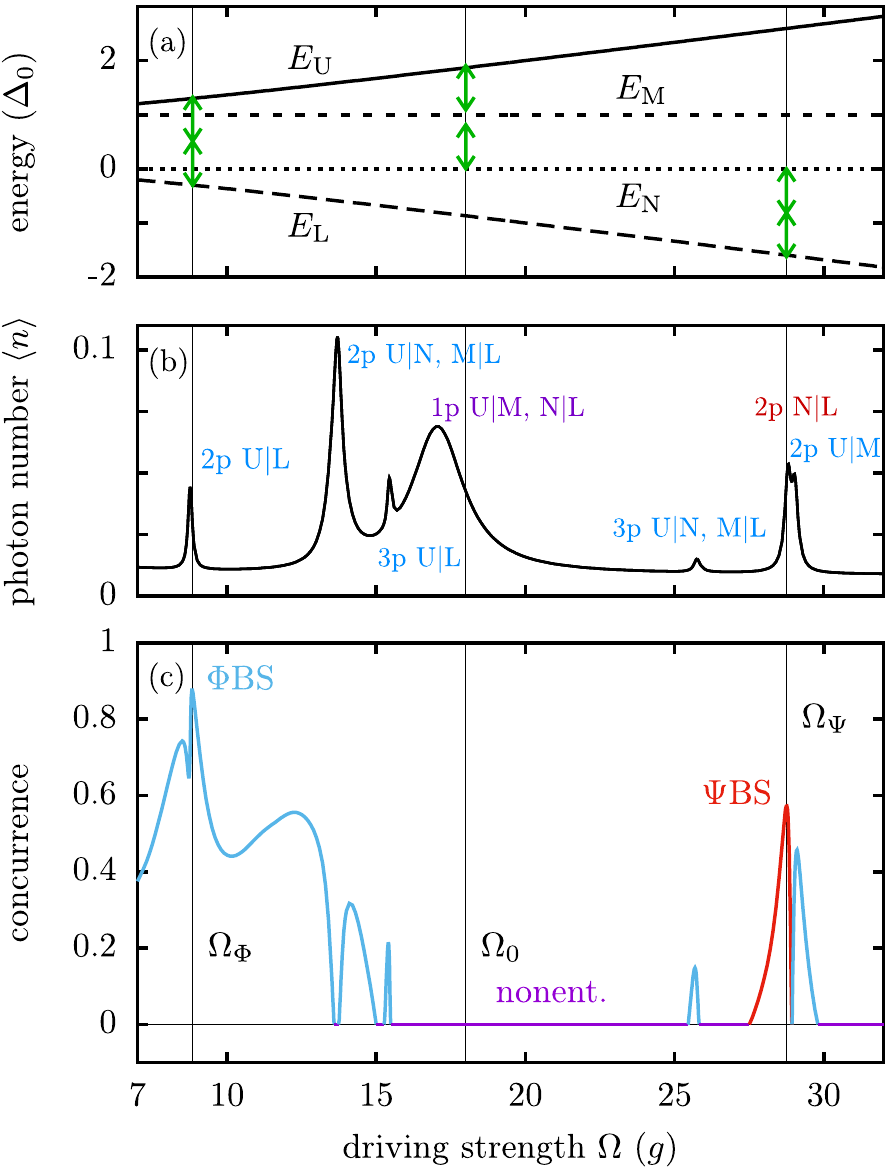}
\caption{
(a) Energies of the four laser-dressed states as function of $\Omega$ (in units of $g$). Green double-headed arrows symbolize the cavity mode energy. (b) Mean photon number $\langle n\rangle$ and (c) concurrence as functions of the driving strength $\Omega$ for a cavity laser detuning $\Delta=0.8\Delta_0$. $n$-photon resonances between the dressed states $\KET{\chi_1}$ and $\KET{\chi_2}$ are labeled by $n$p $\chi_1\vert\chi_2$. The type of entanglement is color-coded: blue = $\Phi$BS entanglement, red = $\Psi$BS entanglement, purple = no entanglement. Straight lines mark the driving strengths used for switching in Fig.~\ref{fig:switching}. 
}
\label{fig:con_vs_Omega}
\end{figure}

Figure~\ref{fig:con_vs_Omega} depicts the dressed state energies [panel (a)], the mean photon number $\n = \langle\aHd\aH+\aVd\aV\rangle$ [panel (b)], and the concurrence [(panel (c)] as functions of the driving strength $\Omega$. All quantities are calculated at times where the system has reached its steady state, i.e., it is assumed that the coincidence measurements necessary to determine $\rho^\text{2p}$ and $C$ are performed after the steady state in the system dynamics has been achieved\cite{Note3}.
\footnotetext[3]{Note that in the steady state situation with constant continuous driving, $\rho^\text{2p}$ and $C$ do not depend on $\Delta t$ since the statistical operator and the Hamiltonian $\hat{H}$ are constant during the measurement process.}
A color code is used to distinguish between $\Phi$BS (blue) entanglement, $\Psi$BS (red) entanglement, and nonentangled photon pairs (purple).

The mean photon number exhibits a series of differently shaped peaks related to $n$-photon transitions between the four laser-dressed states. An $n$-photon transition between a pair of dressed states $\KET{\chi_1}$ and $\KET{\chi_2}$, labeled as $n$p $\chi_1\vert\chi_2$ in Fig.~\ref{fig:con_vs_Omega}(b), is in resonance with the cavity modes when $n$-times the cavity laser detuning $\Delta$ matches the transition energy $E_{\chi_1}-E_{\chi_2}$. Based on this condition, all peaks of enhanced photon production can be linked to one-, two-, or three-photon resonances between the dressed states. In particular, two-photon resonances manifest themselves as high and narrow peaks, e.g., for $\Omega\approx 9g$, $14g$ or $29g$.

Turning to the concurrence, presented in Fig.~\ref{fig:con_vs_Omega}(c), one obtains again a peak-like structure and both types of Bell state entanglement occur. By comparing the concurrences and $\n$, one notes that the regions of high entanglement are associated with two-photon resonances. A more detailed analysis reveals that the features observable for $\Omega\approx 14g$ ($29g$) are actually caused by two closely spaced resonances, 2p U$\vert$N and 2p M$\vert$L (2p U$\vert$M and 2p N$\vert$L), which results in a double peak in the concurrence. A particularly high degree of $\Phi$BS entanglement is obtained for $\Omega_\Phi=8.85g$ when the cavity mode is almost at resonance with the two-photon transition between the dressed states $\KET{U}$ and $\KET{L}$, while at $\Omega_\Psi=28.75g$ a high $\Psi$BS entanglement occurs at the two-photon transition between $\KET{N}$ and $\KET{L}$. This behavior can be well understood using an analysis based on a Schrieffer-Wolff transformation \cite{Seidelmann_QUTE_2020}. Additionally, three-photon resonances lead to small peaks in the concurrence and in the mean photon number.

Besides the regions of high $\Phi$BS and $\Psi$BS entanglement, also a wide regime of vanishing concurrence is found, between $\Omega=16...25 g$, where the cavity modes do not match any multi-photon transition process, cf., Fig.~\ref{fig:con_vs_Omega}. Note that the vanishing degree of entanglement in this parameter regime is not due to a lack of emitted photons. On the contrary, the photon generation can be comparatively high due to the proximity to one-photon resonances, cf., Fig.~\ref{fig:con_vs_Omega}(b). Therefore, in this parameter regime, the measurement detects two subsequent photons that are not entangled. 

According to our findings we choose three driving strengths $\Omega_j$ with similar photon number, but different types of entanglement for the switching protocol: At $\Omega_\Phi=8.85g$ we have a strong $\Phi$BS entanglement, at $\Omega_0=18.00$ we have no entanglement, and at $\Omega_\Psi=28.75g$ we have a strong $\Psi$BS entanglement.

\begin{figure*}[ht]
\centering
\includegraphics[width=\textwidth]{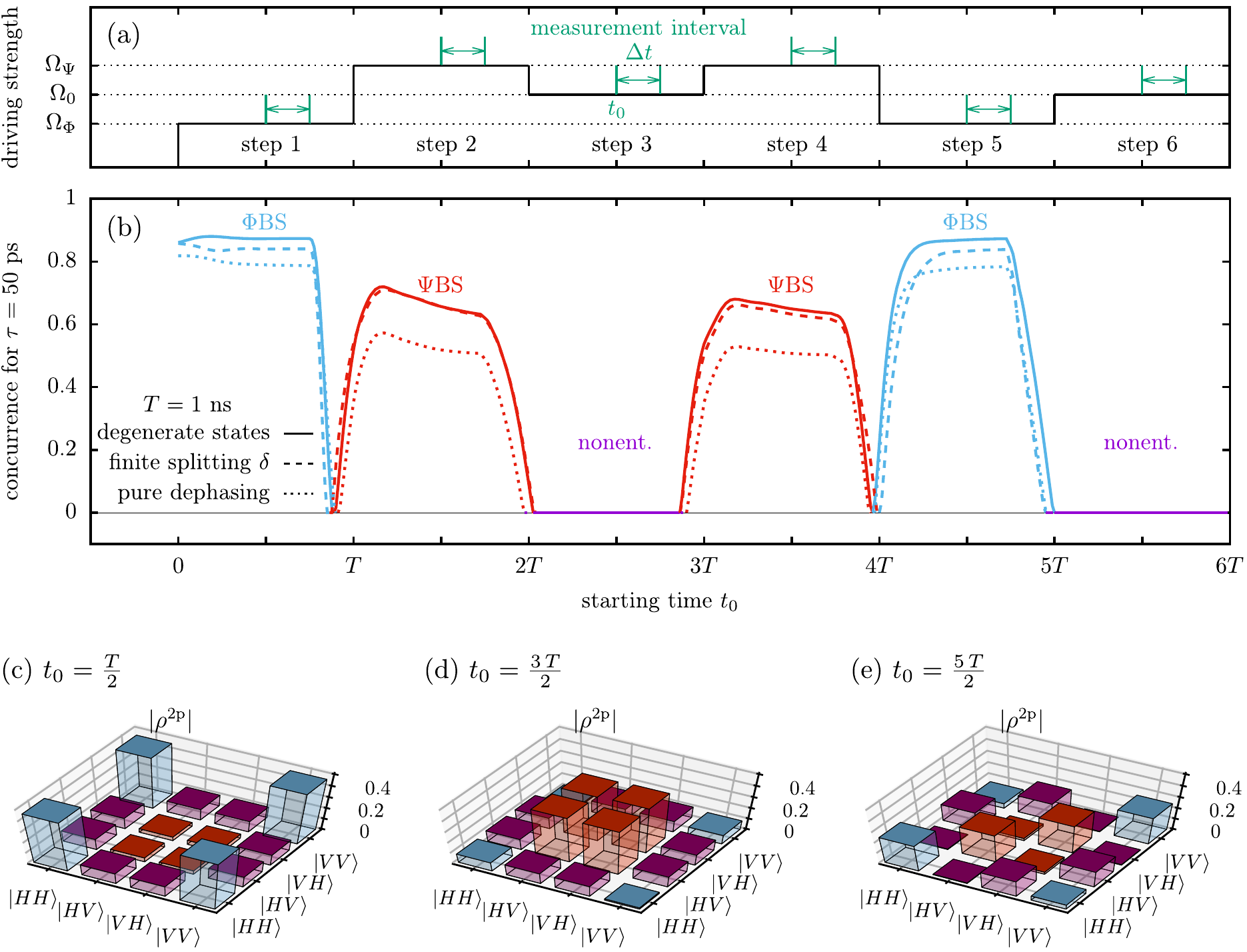}
\caption{
(a) The proposed protocol that enables time-dependent entanglement switching. The driving strength is changed instantaneously between the three values $\Omega_\Phi$, $\Omega_\Psi$, and $\Omega_0$ after a time interval $T$, resulting in a step-like time-dependent laser driving with step length $T$. During each step $j$ coincidence measurements with starting time $t_0$, measurement interval $\Delta t=T/4$, and delay time window $\tau=50$ ps can be performed.
(b) Concurrence calculated for the respective measurements as a function of the starting time $t_0$ for a step length $T=1$ ns. Results are calculated for degenerate intermediate states $\KET{X_\text{H/V}}$ (solid line), for the finite fine-structure splitting $\delta=0.1\Delta_0$ between them (dashed line), and including pure dephasing with $\hbar\gamma_\text{PD}=3$ $\mu$eV (dotted line). The cavity laser detuning is set to $\Delta = 0.8\Delta_0$ and the driving strength values $\Omega_\Phi=8.85g$, $\Omega_\Psi=28.75g$, and $\Omega_0=18g$ are used. A color code indicates $\Phi$BS (blue) and $\Psi$BS (red) entanglement as well as nonentangled photon pairs (purple).
(c)-(e) Corresponding two-photon density matrices $\rho^\text{2p}$ obtained for the measurements performed at $t_0=T/2$, $3T/2$, and $5T/2$ for the case of degenerate intermediate states.
}
\label{fig:switching}
\end{figure*}

We propose a step-like excitation protocol to demonstrate time-dependent entanglement switching. The results are presented in Fig.~\ref{fig:switching}. A schematic sketch of the protocol is depicted in Fig.~\ref{fig:switching}(a). The basic idea is to change between three different driving strengths $\Omega_j$ that, in the stationary case, are associated with different types of entangled photon pairs. During the protocol, the FLE is continuously driven with a constant driving strength $\Omega_j$ for a fixed time period $T$ and then $\Omega$ changes step-like to one of the other two values. Accordingly, the resulting time-dependent laser driving has a step-like structure with step length $T$. In order to allow for a time resolved detection of the entanglement type, measurements with measurement interval $\Delta t= T/4$, delay time window $\tau=50$ ps, and varying starting times $t_0$ are performed.

Figure~\ref{fig:switching}(b) displays the calculated concurrence for each measurement as a function of its respective starting time $t_0$, where a step length of $T=1$ ns is assumed. As before, the entanglement type is color coded: blue (red) indicates $\Phi$BS ($\Psi$BS) entanglement and purple symbolizes nonentangled photon pairs. The corresponding two-photon density matrices for the measurements performed at $t_0=T/2$, $3T/2$, and $5T/2$ are depicted in Fig.~\ref{fig:switching}(c)-(e). 

The protocol starts with driving strength $\Omega_\Phi$ and indeed $\Phi$BS entanglement with a high concurrence is obtained. The corresponding two-photon density matrix shown in Fig.~\ref{fig:switching}(c) represents a two-photon state close to a maximally entangled $\Phi$BS. We find that the occupations of the states with two equally polarized photons, $\KET{HH}$ and $\KET{VV}$, and the coherence between them dominate $\rho^\text{2p}$ such that their absolute values are close to 1/2. In the second step we switch to $\Omega_\Psi$ and obtain a high concurrence related to $\Psi$BS entanglement. In the two-photon density matrix, presented in Fig.~\ref{fig:switching}(d), the states $\KET{HV}$ and $\KET{VH}$ display the highest occupations and coherence values. In the third step with $\Omega_0$, the entanglement is switched off with zero concurrence. The corresponding, reconstructed density matrix is similar to a statistical mixture, where the coherences needed for an entangled Bell state are practically absent, resulting in a vanishing degree of entanglement.

Having demonstrated that all types of entanglement can be created, we continue the protocol demonstrating that the order of switching does not play a role. Accordingly, we switch in step 4 into $\Psi$BS entanglement, in step 5 we switch into $\Phi$BS entanglement and in step 6 back to no entanglement. The obtained concurrence is similar to that in step 1-3. We also checked that density matrices $\rho^\text{2p}$ obtained in the middle of steps 4, 5 and 6 are almost identical to those presented in  Fig.~\ref{fig:switching}(c)-(e) for the respective driving strength (not shown).

It is also interesting to look at the case when the measurements start in the vicinity of switching times $jT$, where $j\in\lbrace 1,2,...,5\rbrace$. Here, one observes a continuous transition between the different entanglement types. This transition begins when the measurement starting at $t_0$ extends into the next step, i.e., when $t_0\geq jT-\Delta t$. During this transition process the degree of entanglement, as measured by the concurrence, passes through zero when one switches between $\Phi$BS and $\Psi$BS entanglement, or vice versa. After a short transition interval the measured concurrence enters either a plateau of high entanglement associated with the used driving strength or remains zero, when the driving strength is $\Omega_0$.
 
An important question is, how sensitive the proposed protocol is to parameter variations. The main requirement is that different types of entanglement can be obtained at different driving strength values. While regions of high $\Phi$BS entanglement can be found rather easily, $\Psi$BS entanglement occurs not so often. Only the two-photon transition 2p~N$\vert$L always features $\Psi$BS entanglement, while for high driving strengths it can be found also at the 2p~U$\vert$L resonance \cite{Seidelmann_QUTE_2020}. Furthermore, the necessary precondition to obtain $\Psi$BS entanglement at these resonances is a finite detuning $\Delta_0$. In principle, in these situations, one can then switch between the different entanglement types using any finite cavity laser detuning $\Delta$. Hence we expect that the protocol also works for different values of $\Delta_0$ and $\Delta$. However, a more elaborate analysis suggests that high concurrence values for both entanglement types are only obtained if $\Delta$ and $\Delta_0$ are of the same order.

Another possible perturbation is an energy difference between the single-excited states $\KET{X_\text{H/V}}$, which in quantum dots is known as the fine-structure splitting (FSS). A finite FSS, defined as $\delta=\hbar\omega_\mathrm{X_H}-\hbar\omega_\mathrm{X_V}$, between the energies of the intermediate bare states $\KET{X_\text{H/V}}$, is regarded as a main obstacle for entanglement generation \cite{Hafenbrak,Biexc_FSS_electrical_control_Bennett,Bounouar18, Seidelmann2019,Jahnke2012}, because it introduces which-path information and, thus, reduces the degree of entanglement\cite{Jahnke2012,entangled-photon2,Seidelmann2019}. 

To consider the effect of a FSS on the switching protocol and entangled photon pair generation, we included a FSS of $\delta=0.1\Delta_0$ in our calculations [dashed line in Fig.~\ref{fig:switching}(b)], which is a typical value being one order of magnitude smaller than the binding energy\cite{Hafenbrak,Bounouar18,entangled-photon2,entangled-photon1}. We find that this rather large FSS only marginally reduces the concurrence compared with the previous results. The reason is that the transitions in the driven system take place between the laser-dressed states. The FSS affects the energies of the laser-dressed states and their composition only weakly such that the resonance conditions and optical selection rules hold. This implies that the generated photonic states are practically the same and the proposed protocol is robust with respect to a non-zero FSS.

By adding a phenomenological rate model\cite{Jahnke2012,Troiani2006}
\begin{equation}
\mathcal{L}_\text{PD}\hat{\rho} = -\frac{1}{2} \sum\limits_{\substack{\chi,\chi^\prime \\ \chi\neq\chi^\prime}} \gamma_\text{PD} \PRO{\chi}\hat{\rho}\PRO{\chi^\prime}
\end{equation}
with rate $\gamma_\text{PD}$ and $\chi,\chi^\prime\in\lbrace G,X_\text{H},X_\text{V},XX\rbrace$ acting on the statistical operator $\hat{\rho}$, we furthermore consider the influence of pure dephasing. Using a realistic value for quantum dots at low temperatures\cite{Jahnke2012}, $\hbar\gamma_\text{PD}=3$ $\mu$eV, we find that, although the concurrence is reduced, all essential features are unaffected. In particular, one can still switch between different entanglement types with corresponding concurrence $C\geq 0.5$ [dotted line in Fig.~\ref{fig:switching}(b)].


In conclusion, this work presents a protocol for time-dependent entanglement switching based on a driven four-level emitter--cavity system. The protocol is operated by simply switching between different driving strengths in a step-like manner. Depending on the driving strength, one obtains either $\Phi$BS entanglement, $\Psi$BS entanglement or nonentangled photon pairs in the respective measurements. Thus, this work demonstrates a possibility to actively switch between different types of entanglement using a time-dependent external laser excitation. The protocol is also robust against a possible FSS. It is stressed that the protocol enables one to achieve different types of entanglement within the same basis and without further post-processing of the generated photons. 

The proposed protocol is therefore a suitable candidate for the realization of time-dependent entanglement switching which is an important step towards future applications.


%
%

%

\begin{acknowledgments}
D. E. Reiter acknowledges support by the Deutsche Forschungsgemeinschaft (DFG) via the project 428026575.
We are further grateful for support by the Deutsche
Forschungsgemeinschaft (DFG, German Research Foundation) via the project 419036043.
\end{acknowledgments}

\section*{Data availability}
The data that support the findings of this study are available from the corresponding author upon reasonable request.


\begin{thebibliography}{53}%
\makeatletter
\providecommand \@ifxundefined [1]{%
 \@ifx{#1\undefined}
}%
\providecommand \@ifnum [1]{%
 \ifnum #1\expandafter \@firstoftwo
 \else \expandafter \@secondoftwo
 \fi
}%
\providecommand \@ifx [1]{%
 \ifx #1\expandafter \@firstoftwo
 \else \expandafter \@secondoftwo
 \fi
}%
\providecommand \natexlab [1]{#1}%
\providecommand \enquote  [1]{``#1''}%
\providecommand \bibnamefont  [1]{#1}%
\providecommand \bibfnamefont [1]{#1}%
\providecommand \citenamefont [1]{#1}%
\providecommand \href@noop [0]{\@secondoftwo}%
\providecommand \href [0]{\begingroup \@sanitize@url \@href}%
\providecommand \@href[1]{\@@startlink{#1}\@@href}%
\providecommand \@@href[1]{\endgroup#1\@@endlink}%
\providecommand \@sanitize@url [0]{\catcode `\\12\catcode `\$12\catcode
  `\&12\catcode `\#12\catcode `\^12\catcode `\_12\catcode `\%12\relax}%
\providecommand \@@startlink[1]{}%
\providecommand \@@endlink[0]{}%
\providecommand \url  [0]{\begingroup\@sanitize@url \@url }%
\providecommand \@url [1]{\endgroup\@href {#1}{\urlprefix }}%
\providecommand \urlprefix  [0]{URL }%
\providecommand \Eprint [0]{\href }%
\providecommand \doibase [0]{http://dx.doi.org/}%
\providecommand \selectlanguage [0]{\@gobble}%
\providecommand \bibinfo  [0]{\@secondoftwo}%
\providecommand \bibfield  [0]{\@secondoftwo}%
\providecommand \translation [1]{[#1]}%
\providecommand \BibitemOpen [0]{}%
\providecommand \bibitemStop [0]{}%
\providecommand \bibitemNoStop [0]{.\EOS\space}%
\providecommand \EOS [0]{\spacefactor3000\relax}%
\providecommand \BibitemShut  [1]{\csname bibitem#1\endcsname}%
\let\auto@bib@innerbib\@empty
\bibitem [{\citenamefont {Gisin}\ \emph {et~al.}(2002)\citenamefont {Gisin},
  \citenamefont {Ribordy}, \citenamefont {Tittel},\ and\ \citenamefont
  {Zbinden}}]{Gisin:02}%
  \BibitemOpen
  \bibfield  {author} {\bibinfo {author} {\bibfnamefont {N.}~\bibnamefont
  {Gisin}}, \bibinfo {author} {\bibfnamefont {G.}~\bibnamefont {Ribordy}},
  \bibinfo {author} {\bibfnamefont {W.}~\bibnamefont {Tittel}}, \ and\ \bibinfo
  {author} {\bibfnamefont {H.}~\bibnamefont {Zbinden}},\ }\href {\doibase
  10.1103/RevModPhys.74.145} {\bibfield  {journal} {\bibinfo  {journal} {Rev.
  Mod. Phys.}\ }\textbf {\bibinfo {volume} {74}},\ \bibinfo {pages} {145}
  (\bibinfo {year} {2002})}\BibitemShut {NoStop}%
\bibitem [{\citenamefont {Lo}, \citenamefont {Curty},\ and\ \citenamefont
  {Tamaki}(2014)}]{Lo_quantum_cryptography}%
  \BibitemOpen
  \bibfield  {author} {\bibinfo {author} {\bibfnamefont {H.-K.}\ \bibnamefont
  {Lo}}, \bibinfo {author} {\bibfnamefont {M.}~\bibnamefont {Curty}}, \ and\
  \bibinfo {author} {\bibfnamefont {K.}~\bibnamefont {Tamaki}},\ }\href
  {\doibase 10.1038/nphoton.2014.149} {\bibfield  {journal} {\bibinfo
  {journal} {Nat. Photonics}\ }\textbf {\bibinfo {volume} {8}},\ \bibinfo
  {pages} {595} (\bibinfo {year} {2014})}\BibitemShut {NoStop}%
\bibitem [{\citenamefont {Duan}\ \emph {et~al.}(2001)\citenamefont {Duan},
  \citenamefont {Lukin}, \citenamefont {Cirac},\ and\ \citenamefont
  {Zoller}}]{duan_quantum_comm}%
  \BibitemOpen
  \bibfield  {author} {\bibinfo {author} {\bibfnamefont {L.-M.}\ \bibnamefont
  {Duan}}, \bibinfo {author} {\bibfnamefont {M.~D.}\ \bibnamefont {Lukin}},
  \bibinfo {author} {\bibfnamefont {J.~I.}\ \bibnamefont {Cirac}}, \ and\
  \bibinfo {author} {\bibfnamefont {P.}~\bibnamefont {Zoller}},\ }\href
  {\doibase 10.1038/35106500} {\bibfield  {journal} {\bibinfo  {journal}
  {Nature}\ }\textbf {\bibinfo {volume} {414}},\ \bibinfo {pages} {413}
  (\bibinfo {year} {2001})}\BibitemShut {NoStop}%
\bibitem [{\citenamefont {Huber}\ \emph
  {et~al.}(2018{\natexlab{a}})\citenamefont {Huber}, \citenamefont {Reindl},
  \citenamefont {Aberl}, \citenamefont {Rastelli},\ and\ \citenamefont
  {Trotta}}]{Huber_overview_2018}%
  \BibitemOpen
  \bibfield  {author} {\bibinfo {author} {\bibfnamefont {D.}~\bibnamefont
  {Huber}}, \bibinfo {author} {\bibfnamefont {M.}~\bibnamefont {Reindl}},
  \bibinfo {author} {\bibfnamefont {J.}~\bibnamefont {Aberl}}, \bibinfo
  {author} {\bibfnamefont {A.}~\bibnamefont {Rastelli}}, \ and\ \bibinfo
  {author} {\bibfnamefont {R.}~\bibnamefont {Trotta}},\ }\href {\doibase
  10.1088/2040-8986/aac4c4} {\bibfield  {journal} {\bibinfo  {journal} {Journal
  of Optics}\ }\textbf {\bibinfo {volume} {20}},\ \bibinfo {pages} {073002}
  (\bibinfo {year} {2018}{\natexlab{a}})}\BibitemShut {NoStop}%
\bibitem [{\citenamefont {Pan}\ \emph {et~al.}(2012)\citenamefont {Pan},
  \citenamefont {Chen}, \citenamefont {Lu}, \citenamefont {Weinfurter},
  \citenamefont {Zeilinger},\ and\ \citenamefont {\ifmmode~\dot{Z}\else
  \.{Z}\fi{}ukowski}}]{pan:12}%
  \BibitemOpen
  \bibfield  {author} {\bibinfo {author} {\bibfnamefont {J.-W.}\ \bibnamefont
  {Pan}}, \bibinfo {author} {\bibfnamefont {Z.-B.}\ \bibnamefont {Chen}},
  \bibinfo {author} {\bibfnamefont {C.-Y.}\ \bibnamefont {Lu}}, \bibinfo
  {author} {\bibfnamefont {H.}~\bibnamefont {Weinfurter}}, \bibinfo {author}
  {\bibfnamefont {A.}~\bibnamefont {Zeilinger}}, \ and\ \bibinfo {author}
  {\bibfnamefont {M.}~\bibnamefont {\ifmmode~\dot{Z}\else \.{Z}\fi{}ukowski}},\
  }\href {\doibase 10.1103/RevModPhys.84.777} {\bibfield  {journal} {\bibinfo
  {journal} {Rev. Mod. Phys.}\ }\textbf {\bibinfo {volume} {84}},\ \bibinfo
  {pages} {777} (\bibinfo {year} {2012})}\BibitemShut {NoStop}%
\bibitem [{\citenamefont {Bennett}\ and\ \citenamefont
  {DiVincenzo}(2000)}]{Bennett:00}%
  \BibitemOpen
  \bibfield  {author} {\bibinfo {author} {\bibfnamefont {C.~H.}\ \bibnamefont
  {Bennett}}\ and\ \bibinfo {author} {\bibfnamefont {D.~P.}\ \bibnamefont
  {DiVincenzo}},\ }\href {https://doi.org/10.1038/35005001} {\bibfield
  {journal} {\bibinfo  {journal} {Nature}\ }\textbf {\bibinfo {volume} {404}},\
  \bibinfo {pages} {247} (\bibinfo {year} {2000})}\BibitemShut {NoStop}%
\bibitem [{\citenamefont {Kuhn}\ \emph {et~al.}(2016)\citenamefont {Kuhn},
  \citenamefont {Knorr}, \citenamefont {Reitzenstein},\ and\ \citenamefont
  {Richter}}]{Kuhn:16}%
  \BibitemOpen
  \bibfield  {author} {\bibinfo {author} {\bibfnamefont {S.~C.}\ \bibnamefont
  {Kuhn}}, \bibinfo {author} {\bibfnamefont {A.}~\bibnamefont {Knorr}},
  \bibinfo {author} {\bibfnamefont {S.}~\bibnamefont {Reitzenstein}}, \ and\
  \bibinfo {author} {\bibfnamefont {M.}~\bibnamefont {Richter}},\ }\href
  {\doibase 10.1364/OE.24.025446} {\bibfield  {journal} {\bibinfo  {journal}
  {Opt. Express}\ }\textbf {\bibinfo {volume} {24}},\ \bibinfo {pages} {25446}
  (\bibinfo {year} {2016})}\BibitemShut {NoStop}%
\bibitem [{\citenamefont {Zeilinger}(2017)}]{Zeilinger_entangled}%
  \BibitemOpen
  \bibfield  {author} {\bibinfo {author} {\bibfnamefont {A.}~\bibnamefont
  {Zeilinger}},\ }\href {http://stacks.iop.org/1402-4896/92/i=7/a=072501}
  {\bibfield  {journal} {\bibinfo  {journal} {Phys. Scr.}\ }\textbf {\bibinfo
  {volume} {92}},\ \bibinfo {pages} {072501} (\bibinfo {year}
  {2017})}\BibitemShut {NoStop}%
\bibitem [{\citenamefont {Orieux}\ \emph {et~al.}(2017)\citenamefont {Orieux},
  \citenamefont {Versteegh}, \citenamefont {J{\"o}ns},\ and\ \citenamefont
  {Ducci}}]{Orieux_entangled}%
  \BibitemOpen
  \bibfield  {author} {\bibinfo {author} {\bibfnamefont {A.}~\bibnamefont
  {Orieux}}, \bibinfo {author} {\bibfnamefont {M.~A.~M.}\ \bibnamefont
  {Versteegh}}, \bibinfo {author} {\bibfnamefont {K.~D.}\ \bibnamefont
  {J{\"o}ns}}, \ and\ \bibinfo {author} {\bibfnamefont {S.}~\bibnamefont
  {Ducci}},\ }\href {http://stacks.iop.org/0034-4885/80/i=7/a=076001}
  {\bibfield  {journal} {\bibinfo  {journal} {Rep. Prog. Phys.}\ }\textbf
  {\bibinfo {volume} {80}},\ \bibinfo {pages} {076001} (\bibinfo {year}
  {2017})}\BibitemShut {NoStop}%
\bibitem [{\citenamefont {Edamatsu}(2007)}]{edamatsu2007entangled}%
  \BibitemOpen
  \bibfield  {author} {\bibinfo {author} {\bibfnamefont {K.}~\bibnamefont
  {Edamatsu}},\ }\href {\doibase 10.1143/JJAP.46.7175} {\bibfield  {journal}
  {\bibinfo  {journal} {Jpn. J. Appl. Phys.}\ }\textbf {\bibinfo {volume}
  {46}},\ \bibinfo {pages} {7175} (\bibinfo {year} {2007})}\BibitemShut
  {NoStop}%
\bibitem [{\citenamefont {Freedman}\ and\ \citenamefont
  {Clauser}(1972)}]{Freedman_FLE_atom}%
  \BibitemOpen
  \bibfield  {author} {\bibinfo {author} {\bibfnamefont {S.~J.}\ \bibnamefont
  {Freedman}}\ and\ \bibinfo {author} {\bibfnamefont {J.~F.}\ \bibnamefont
  {Clauser}},\ }\href {\doibase 10.1103/PhysRevLett.28.938} {\bibfield
  {journal} {\bibinfo  {journal} {Phys. Rev. Lett.}\ }\textbf {\bibinfo
  {volume} {28}},\ \bibinfo {pages} {938} (\bibinfo {year} {1972})}\BibitemShut
  {NoStop}%
\bibitem [{\citenamefont {Wen}\ \emph {et~al.}(2008)\citenamefont {Wen},
  \citenamefont {Du}, \citenamefont {Zhang}, \citenamefont {Xiao},\ and\
  \citenamefont {Rubin}}]{Wen_theory_atomic_system}%
  \BibitemOpen
  \bibfield  {author} {\bibinfo {author} {\bibfnamefont {J.}~\bibnamefont
  {Wen}}, \bibinfo {author} {\bibfnamefont {S.}~\bibnamefont {Du}}, \bibinfo
  {author} {\bibfnamefont {Y.}~\bibnamefont {Zhang}}, \bibinfo {author}
  {\bibfnamefont {M.}~\bibnamefont {Xiao}}, \ and\ \bibinfo {author}
  {\bibfnamefont {M.~H.}\ \bibnamefont {Rubin}},\ }\href {\doibase
  10.1103/PhysRevA.77.033816} {\bibfield  {journal} {\bibinfo  {journal} {Phys.
  Rev. A}\ }\textbf {\bibinfo {volume} {77}},\ \bibinfo {pages} {033816}
  (\bibinfo {year} {2008})}\BibitemShut {NoStop}%
\bibitem [{\citenamefont {Park}\ \emph {et~al.}(2018)\citenamefont {Park},
  \citenamefont {Jeong}, \citenamefont {Kim},\ and\ \citenamefont
  {Moon}}]{Park_FLE_atom}%
  \BibitemOpen
  \bibfield  {author} {\bibinfo {author} {\bibfnamefont {J.}~\bibnamefont
  {Park}}, \bibinfo {author} {\bibfnamefont {T.}~\bibnamefont {Jeong}},
  \bibinfo {author} {\bibfnamefont {H.}~\bibnamefont {Kim}}, \ and\ \bibinfo
  {author} {\bibfnamefont {H.~S.}\ \bibnamefont {Moon}},\ }\href {\doibase
  10.1103/PhysRevLett.121.263601} {\bibfield  {journal} {\bibinfo  {journal}
  {Phys. Rev. Lett.}\ }\textbf {\bibinfo {volume} {121}},\ \bibinfo {pages}
  {263601} (\bibinfo {year} {2018})}\BibitemShut {NoStop}%
\bibitem [{\citenamefont {Seidelmann}\ \emph
  {et~al.}(2019{\natexlab{a}})\citenamefont {Seidelmann}, \citenamefont
  {Ungar}, \citenamefont {Cygorek}, \citenamefont {Vagov}, \citenamefont
  {Barth}, \citenamefont {Kuhn},\ and\ \citenamefont {Axt}}]{Seidelmann2019}%
  \BibitemOpen
  \bibfield  {author} {\bibinfo {author} {\bibfnamefont {T.}~\bibnamefont
  {Seidelmann}}, \bibinfo {author} {\bibfnamefont {F.}~\bibnamefont {Ungar}},
  \bibinfo {author} {\bibfnamefont {M.}~\bibnamefont {Cygorek}}, \bibinfo
  {author} {\bibfnamefont {A.}~\bibnamefont {Vagov}}, \bibinfo {author}
  {\bibfnamefont {A.~M.}\ \bibnamefont {Barth}}, \bibinfo {author}
  {\bibfnamefont {T.}~\bibnamefont {Kuhn}}, \ and\ \bibinfo {author}
  {\bibfnamefont {V.~M.}\ \bibnamefont {Axt}},\ }\href {\doibase
  10.1103/PhysRevB.99.245301} {\bibfield  {journal} {\bibinfo  {journal} {Phys.
  Rev. B}\ }\textbf {\bibinfo {volume} {99}},\ \bibinfo {pages} {245301}
  (\bibinfo {year} {2019}{\natexlab{a}})}\BibitemShut {NoStop}%
\bibitem [{\citenamefont {Cygorek}\ \emph {et~al.}(2018)\citenamefont
  {Cygorek}, \citenamefont {Ungar}, \citenamefont {Seidelmann}, \citenamefont
  {Barth}, \citenamefont {Vagov}, \citenamefont {Axt},\ and\ \citenamefont
  {Kuhn}}]{Different-Concurrences:18}%
  \BibitemOpen
  \bibfield  {author} {\bibinfo {author} {\bibfnamefont {M.}~\bibnamefont
  {Cygorek}}, \bibinfo {author} {\bibfnamefont {F.}~\bibnamefont {Ungar}},
  \bibinfo {author} {\bibfnamefont {T.}~\bibnamefont {Seidelmann}}, \bibinfo
  {author} {\bibfnamefont {A.~M.}\ \bibnamefont {Barth}}, \bibinfo {author}
  {\bibfnamefont {A.}~\bibnamefont {Vagov}}, \bibinfo {author} {\bibfnamefont
  {V.~M.}\ \bibnamefont {Axt}}, \ and\ \bibinfo {author} {\bibfnamefont
  {T.}~\bibnamefont {Kuhn}},\ }\href {\doibase 10.1103/PhysRevB.98.045303}
  {\bibfield  {journal} {\bibinfo  {journal} {Phys. Rev. B}\ }\textbf {\bibinfo
  {volume} {98}},\ \bibinfo {pages} {045303} (\bibinfo {year}
  {2018})}\BibitemShut {NoStop}%
\bibitem [{\citenamefont {Seidelmann}\ \emph
  {et~al.}(2019{\natexlab{b}})\citenamefont {Seidelmann}, \citenamefont
  {Ungar}, \citenamefont {Barth}, \citenamefont {Vagov}, \citenamefont {Axt},
  \citenamefont {Cygorek},\ and\ \citenamefont
  {Kuhn}}]{Phon_enhanced_entanglement}%
  \BibitemOpen
  \bibfield  {author} {\bibinfo {author} {\bibfnamefont {T.}~\bibnamefont
  {Seidelmann}}, \bibinfo {author} {\bibfnamefont {F.}~\bibnamefont {Ungar}},
  \bibinfo {author} {\bibfnamefont {A.~M.}\ \bibnamefont {Barth}}, \bibinfo
  {author} {\bibfnamefont {A.}~\bibnamefont {Vagov}}, \bibinfo {author}
  {\bibfnamefont {V.~M.}\ \bibnamefont {Axt}}, \bibinfo {author} {\bibfnamefont
  {M.}~\bibnamefont {Cygorek}}, \ and\ \bibinfo {author} {\bibfnamefont
  {T.}~\bibnamefont {Kuhn}},\ }\href {\doibase 10.1103/PhysRevLett.123.137401}
  {\bibfield  {journal} {\bibinfo  {journal} {Phys. Rev. Lett.}\ }\textbf
  {\bibinfo {volume} {123}},\ \bibinfo {pages} {137401} (\bibinfo {year}
  {2019}{\natexlab{b}})}\BibitemShut {NoStop}%
\bibitem [{\citenamefont {Schumacher}\ \emph {et~al.}(2012)\citenamefont
  {Schumacher}, \citenamefont {F\"{o}rstner}, \citenamefont {Zrenner},
  \citenamefont {Florian}, \citenamefont {Gies}, \citenamefont {Gartner},\ and\
  \citenamefont {Jahnke}}]{Jahnke2012}%
  \BibitemOpen
  \bibfield  {author} {\bibinfo {author} {\bibfnamefont {S.}~\bibnamefont
  {Schumacher}}, \bibinfo {author} {\bibfnamefont {J.}~\bibnamefont
  {F\"{o}rstner}}, \bibinfo {author} {\bibfnamefont {A.}~\bibnamefont
  {Zrenner}}, \bibinfo {author} {\bibfnamefont {M.}~\bibnamefont {Florian}},
  \bibinfo {author} {\bibfnamefont {C.}~\bibnamefont {Gies}}, \bibinfo {author}
  {\bibfnamefont {P.}~\bibnamefont {Gartner}}, \ and\ \bibinfo {author}
  {\bibfnamefont {F.}~\bibnamefont {Jahnke}},\ }\href {\doibase
  10.1364/OE.20.005335} {\bibfield  {journal} {\bibinfo  {journal} {Opt.
  Express}\ }\textbf {\bibinfo {volume} {20}},\ \bibinfo {pages} {5335}
  (\bibinfo {year} {2012})}\BibitemShut {NoStop}%
\bibitem [{\citenamefont {Heinze}, \citenamefont {Zrenner},\ and\ \citenamefont
  {Schumacher}(2017)}]{heinze17}%
  \BibitemOpen
  \bibfield  {author} {\bibinfo {author} {\bibfnamefont {D.}~\bibnamefont
  {Heinze}}, \bibinfo {author} {\bibfnamefont {A.}~\bibnamefont {Zrenner}}, \
  and\ \bibinfo {author} {\bibfnamefont {S.}~\bibnamefont {Schumacher}},\
  }\href {\doibase 10.1103/PhysRevB.95.245306} {\bibfield  {journal} {\bibinfo
  {journal} {Phys. Rev. B}\ }\textbf {\bibinfo {volume} {95}},\ \bibinfo
  {pages} {245306} (\bibinfo {year} {2017})}\BibitemShut {NoStop}%
\bibitem [{\citenamefont {Carmele}\ and\ \citenamefont
  {Knorr}(2011)}]{BiexcCasc_Carmele}%
  \BibitemOpen
  \bibfield  {author} {\bibinfo {author} {\bibfnamefont {A.}~\bibnamefont
  {Carmele}}\ and\ \bibinfo {author} {\bibfnamefont {A.}~\bibnamefont
  {Knorr}},\ }\href {\doibase 10.1103/PhysRevB.84.075328} {\bibfield  {journal}
  {\bibinfo  {journal} {Phys. Rev. B}\ }\textbf {\bibinfo {volume} {84}},\
  \bibinfo {pages} {075328} (\bibinfo {year} {2011})}\BibitemShut {NoStop}%
\bibitem [{\citenamefont {Stevenson}\ \emph {et~al.}(2006)\citenamefont
  {Stevenson}, \citenamefont {Young}, \citenamefont {Atkinson}, \citenamefont
  {Cooper}, \citenamefont {Ritchie},\ and\ \citenamefont
  {Shields}}]{Stevenson2006}%
  \BibitemOpen
  \bibfield  {author} {\bibinfo {author} {\bibfnamefont {R.~M.}\ \bibnamefont
  {Stevenson}}, \bibinfo {author} {\bibfnamefont {R.~J.}\ \bibnamefont
  {Young}}, \bibinfo {author} {\bibfnamefont {P.}~\bibnamefont {Atkinson}},
  \bibinfo {author} {\bibfnamefont {K.}~\bibnamefont {Cooper}}, \bibinfo
  {author} {\bibfnamefont {D.~A.}\ \bibnamefont {Ritchie}}, \ and\ \bibinfo
  {author} {\bibfnamefont {A.~J.}\ \bibnamefont {Shields}},\ }\href
  {http://dx.doi.org/10.1038/nature04446} {\bibfield  {journal} {\bibinfo
  {journal} {Nature}\ }\textbf {\bibinfo {volume} {439}},\ \bibinfo {pages}
  {179} (\bibinfo {year} {2006})}\BibitemShut {NoStop}%
\bibitem [{\citenamefont {Young}\ \emph {et~al.}(2006)\citenamefont {Young},
  \citenamefont {Stevenson}, \citenamefont {Atkinson}, \citenamefont {Cooper},
  \citenamefont {Ritchie},\ and\ \citenamefont {Shields}}]{Young_2006}%
  \BibitemOpen
  \bibfield  {author} {\bibinfo {author} {\bibfnamefont {R.~J.}\ \bibnamefont
  {Young}}, \bibinfo {author} {\bibfnamefont {R.~M.}\ \bibnamefont
  {Stevenson}}, \bibinfo {author} {\bibfnamefont {P.}~\bibnamefont {Atkinson}},
  \bibinfo {author} {\bibfnamefont {K.}~\bibnamefont {Cooper}}, \bibinfo
  {author} {\bibfnamefont {D.~A.}\ \bibnamefont {Ritchie}}, \ and\ \bibinfo
  {author} {\bibfnamefont {A.~J.}\ \bibnamefont {Shields}},\ }\href {\doibase
  10.1088/1367-2630/8/2/029} {\bibfield  {journal} {\bibinfo  {journal} {New J.
  Phys.}\ }\textbf {\bibinfo {volume} {8}},\ \bibinfo {pages} {29} (\bibinfo
  {year} {2006})}\BibitemShut {NoStop}%
\bibitem [{\citenamefont {Muller}\ \emph {et~al.}(2009)\citenamefont {Muller},
  \citenamefont {Fang}, \citenamefont {Lawall},\ and\ \citenamefont
  {Solomon}}]{Muller_2009}%
  \BibitemOpen
  \bibfield  {author} {\bibinfo {author} {\bibfnamefont {A.}~\bibnamefont
  {Muller}}, \bibinfo {author} {\bibfnamefont {W.}~\bibnamefont {Fang}},
  \bibinfo {author} {\bibfnamefont {J.}~\bibnamefont {Lawall}}, \ and\ \bibinfo
  {author} {\bibfnamefont {G.~S.}\ \bibnamefont {Solomon}},\ }\href {\doibase
  10.1103/PhysRevLett.103.217402} {\bibfield  {journal} {\bibinfo  {journal}
  {Phys. Rev. Lett.}\ }\textbf {\bibinfo {volume} {103}},\ \bibinfo {pages}
  {217402} (\bibinfo {year} {2009})}\BibitemShut {NoStop}%
\bibitem [{\citenamefont {Huber}\ \emph
  {et~al.}(2018{\natexlab{b}})\citenamefont {Huber}, \citenamefont {Reindl},
  \citenamefont {Covre~da Silva}, \citenamefont {Schimpf}, \citenamefont
  {Mart\'{\i}n-S\'anchez}, \citenamefont {Huang}, \citenamefont {Piredda},
  \citenamefont {Edlinger}, \citenamefont {Rastelli},\ and\ \citenamefont
  {Trotta}}]{Huber_PRL_2018}%
  \BibitemOpen
  \bibfield  {author} {\bibinfo {author} {\bibfnamefont {D.}~\bibnamefont
  {Huber}}, \bibinfo {author} {\bibfnamefont {M.}~\bibnamefont {Reindl}},
  \bibinfo {author} {\bibfnamefont {S.~F.}\ \bibnamefont {Covre~da Silva}},
  \bibinfo {author} {\bibfnamefont {C.}~\bibnamefont {Schimpf}}, \bibinfo
  {author} {\bibfnamefont {J.}~\bibnamefont {Mart\'{\i}n-S\'anchez}}, \bibinfo
  {author} {\bibfnamefont {H.}~\bibnamefont {Huang}}, \bibinfo {author}
  {\bibfnamefont {G.}~\bibnamefont {Piredda}}, \bibinfo {author} {\bibfnamefont
  {J.}~\bibnamefont {Edlinger}}, \bibinfo {author} {\bibfnamefont
  {A.}~\bibnamefont {Rastelli}}, \ and\ \bibinfo {author} {\bibfnamefont
  {R.}~\bibnamefont {Trotta}},\ }\href {\doibase
  10.1103/PhysRevLett.121.033902} {\bibfield  {journal} {\bibinfo  {journal}
  {Phys. Rev. Lett.}\ }\textbf {\bibinfo {volume} {121}},\ \bibinfo {pages}
  {033902} (\bibinfo {year} {2018}{\natexlab{b}})}\BibitemShut {NoStop}%
\bibitem [{\citenamefont {Wang}\ \emph {et~al.}(2019)\citenamefont {Wang},
  \citenamefont {Hu}, \citenamefont {Chung}, \citenamefont {Qin}, \citenamefont
  {Yang}, \citenamefont {Li}, \citenamefont {Liu}, \citenamefont {Zhong},
  \citenamefont {He}, \citenamefont {Ding}, \citenamefont {Deng}, \citenamefont
  {Dai}, \citenamefont {Huo}, \citenamefont {H\"ofling}, \citenamefont {Lu},\
  and\ \citenamefont {Pan}}]{Wang_2019}%
  \BibitemOpen
  \bibfield  {author} {\bibinfo {author} {\bibfnamefont {H.}~\bibnamefont
  {Wang}}, \bibinfo {author} {\bibfnamefont {H.}~\bibnamefont {Hu}}, \bibinfo
  {author} {\bibfnamefont {T.-H.}\ \bibnamefont {Chung}}, \bibinfo {author}
  {\bibfnamefont {J.}~\bibnamefont {Qin}}, \bibinfo {author} {\bibfnamefont
  {X.}~\bibnamefont {Yang}}, \bibinfo {author} {\bibfnamefont {J.-P.}\
  \bibnamefont {Li}}, \bibinfo {author} {\bibfnamefont {R.-Z.}\ \bibnamefont
  {Liu}}, \bibinfo {author} {\bibfnamefont {H.-S.}\ \bibnamefont {Zhong}},
  \bibinfo {author} {\bibfnamefont {Y.-M.}\ \bibnamefont {He}}, \bibinfo
  {author} {\bibfnamefont {X.}~\bibnamefont {Ding}}, \bibinfo {author}
  {\bibfnamefont {Y.-H.}\ \bibnamefont {Deng}}, \bibinfo {author}
  {\bibfnamefont {Q.}~\bibnamefont {Dai}}, \bibinfo {author} {\bibfnamefont
  {Y.-H.}\ \bibnamefont {Huo}}, \bibinfo {author} {\bibfnamefont
  {S.}~\bibnamefont {H\"ofling}}, \bibinfo {author} {\bibfnamefont {C.-Y.}\
  \bibnamefont {Lu}}, \ and\ \bibinfo {author} {\bibfnamefont {J.-W.}\
  \bibnamefont {Pan}},\ }\href {\doibase 10.1103/PhysRevLett.122.113602}
  {\bibfield  {journal} {\bibinfo  {journal} {Phys. Rev. Lett.}\ }\textbf
  {\bibinfo {volume} {122}},\ \bibinfo {pages} {113602} (\bibinfo {year}
  {2019})}\BibitemShut {NoStop}%
\bibitem [{\citenamefont {Liu}\ \emph {et~al.}(2019)\citenamefont {Liu},
  \citenamefont {Su}, \citenamefont {Wei}, \citenamefont {Yao}, \citenamefont
  {Silva}, \citenamefont {Yu}, \citenamefont {Iles-Smith}, \citenamefont
  {Srinivasan}, \citenamefont {Rastelli}, \citenamefont {Li},\ and\
  \citenamefont {Wang}}]{Liu2019}%
  \BibitemOpen
  \bibfield  {author} {\bibinfo {author} {\bibfnamefont {J.}~\bibnamefont
  {Liu}}, \bibinfo {author} {\bibfnamefont {R.}~\bibnamefont {Su}}, \bibinfo
  {author} {\bibfnamefont {Y.}~\bibnamefont {Wei}}, \bibinfo {author}
  {\bibfnamefont {B.}~\bibnamefont {Yao}}, \bibinfo {author} {\bibfnamefont
  {S.~F. C.~d.}\ \bibnamefont {Silva}}, \bibinfo {author} {\bibfnamefont
  {Y.}~\bibnamefont {Yu}}, \bibinfo {author} {\bibfnamefont {J.}~\bibnamefont
  {Iles-Smith}}, \bibinfo {author} {\bibfnamefont {K.}~\bibnamefont
  {Srinivasan}}, \bibinfo {author} {\bibfnamefont {A.}~\bibnamefont
  {Rastelli}}, \bibinfo {author} {\bibfnamefont {J.}~\bibnamefont {Li}}, \ and\
  \bibinfo {author} {\bibfnamefont {X.}~\bibnamefont {Wang}},\ }\href {\doibase
  10.1038/s41565-019-0435-9} {\bibfield  {journal} {\bibinfo  {journal} {Nat.
  Nanotechnol.}\ }\textbf {\bibinfo {volume} {14}},\ \bibinfo {pages} {586}
  (\bibinfo {year} {2019})}\BibitemShut {NoStop}%
\bibitem [{\citenamefont {Bounouar}\ \emph {et~al.}(2018)\citenamefont
  {Bounouar}, \citenamefont {de~la Haye}, \citenamefont {Strauß},
  \citenamefont {Schnauber}, \citenamefont {Thoma}, \citenamefont {Gschrey},
  \citenamefont {Schulze}, \citenamefont {Strittmatter}, \citenamefont {Rodt},\
  and\ \citenamefont {Reitzenstein}}]{Bounouar18}%
  \BibitemOpen
  \bibfield  {author} {\bibinfo {author} {\bibfnamefont {S.}~\bibnamefont
  {Bounouar}}, \bibinfo {author} {\bibfnamefont {C.}~\bibnamefont {de~la
  Haye}}, \bibinfo {author} {\bibfnamefont {M.}~\bibnamefont {Strauß}},
  \bibinfo {author} {\bibfnamefont {P.}~\bibnamefont {Schnauber}}, \bibinfo
  {author} {\bibfnamefont {A.}~\bibnamefont {Thoma}}, \bibinfo {author}
  {\bibfnamefont {M.}~\bibnamefont {Gschrey}}, \bibinfo {author} {\bibfnamefont
  {J.-H.}\ \bibnamefont {Schulze}}, \bibinfo {author} {\bibfnamefont
  {A.}~\bibnamefont {Strittmatter}}, \bibinfo {author} {\bibfnamefont
  {S.}~\bibnamefont {Rodt}}, \ and\ \bibinfo {author} {\bibfnamefont
  {S.}~\bibnamefont {Reitzenstein}},\ }\href {\doibase 10.1063/1.5020242}
  {\bibfield  {journal} {\bibinfo  {journal} {Appl. Phys. Lett.}\ }\textbf
  {\bibinfo {volume} {112}},\ \bibinfo {pages} {153107} (\bibinfo {year}
  {2018})}\BibitemShut {NoStop}%
\bibitem [{\citenamefont {Dousse}\ \emph {et~al.}(2010)\citenamefont {Dousse},
  \citenamefont {Suffczy{\'n}ski}, \citenamefont {Beveratos}, \citenamefont
  {Krebs}, \citenamefont {Lema{\^i}tre}, \citenamefont {Sagnes}, \citenamefont
  {Bloch}, \citenamefont {Voisin},\ and\ \citenamefont
  {Senellart}}]{dousse:10}%
  \BibitemOpen
  \bibfield  {author} {\bibinfo {author} {\bibfnamefont {A.}~\bibnamefont
  {Dousse}}, \bibinfo {author} {\bibfnamefont {J.}~\bibnamefont
  {Suffczy{\'n}ski}}, \bibinfo {author} {\bibfnamefont {A.}~\bibnamefont
  {Beveratos}}, \bibinfo {author} {\bibfnamefont {O.}~\bibnamefont {Krebs}},
  \bibinfo {author} {\bibfnamefont {A.}~\bibnamefont {Lema{\^i}tre}}, \bibinfo
  {author} {\bibfnamefont {I.}~\bibnamefont {Sagnes}}, \bibinfo {author}
  {\bibfnamefont {J.}~\bibnamefont {Bloch}}, \bibinfo {author} {\bibfnamefont
  {P.}~\bibnamefont {Voisin}}, \ and\ \bibinfo {author} {\bibfnamefont
  {P.}~\bibnamefont {Senellart}},\ }\href {https://doi.org/10.1038/nature09148}
  {\bibfield  {journal} {\bibinfo  {journal} {Nature}\ }\textbf {\bibinfo
  {volume} {466}},\ \bibinfo {pages} {217} (\bibinfo {year}
  {2010})}\BibitemShut {NoStop}%
\bibitem [{\citenamefont {Winik}\ \emph {et~al.}(2017)\citenamefont {Winik},
  \citenamefont {Cogan}, \citenamefont {Don}, \citenamefont {Schwartz},
  \citenamefont {Gantz}, \citenamefont {Schmidgall}, \citenamefont {Livneh},
  \citenamefont {Rapaport}, \citenamefont {Buks},\ and\ \citenamefont
  {Gershoni}}]{winik:2017}%
  \BibitemOpen
  \bibfield  {author} {\bibinfo {author} {\bibfnamefont {R.}~\bibnamefont
  {Winik}}, \bibinfo {author} {\bibfnamefont {D.}~\bibnamefont {Cogan}},
  \bibinfo {author} {\bibfnamefont {Y.}~\bibnamefont {Don}}, \bibinfo {author}
  {\bibfnamefont {I.}~\bibnamefont {Schwartz}}, \bibinfo {author}
  {\bibfnamefont {L.}~\bibnamefont {Gantz}}, \bibinfo {author} {\bibfnamefont
  {E.~R.}\ \bibnamefont {Schmidgall}}, \bibinfo {author} {\bibfnamefont
  {N.}~\bibnamefont {Livneh}}, \bibinfo {author} {\bibfnamefont
  {R.}~\bibnamefont {Rapaport}}, \bibinfo {author} {\bibfnamefont
  {E.}~\bibnamefont {Buks}}, \ and\ \bibinfo {author} {\bibfnamefont
  {D.}~\bibnamefont {Gershoni}},\ }\href {\doibase 10.1103/PhysRevB.95.235435}
  {\bibfield  {journal} {\bibinfo  {journal} {Phys. Rev. B}\ }\textbf {\bibinfo
  {volume} {95}},\ \bibinfo {pages} {235435} (\bibinfo {year}
  {2017})}\BibitemShut {NoStop}%
\bibitem [{\citenamefont {M{\"u}ller}\ \emph {et~al.}(2014)\citenamefont
  {M{\"u}ller}, \citenamefont {Bounouar}, \citenamefont {J{\"o}ns},
  \citenamefont {Gl{\"a}ssl},\ and\ \citenamefont
  {Michler}}]{entangled-photon1}%
  \BibitemOpen
  \bibfield  {author} {\bibinfo {author} {\bibfnamefont {M.}~\bibnamefont
  {M{\"u}ller}}, \bibinfo {author} {\bibfnamefont {S.}~\bibnamefont
  {Bounouar}}, \bibinfo {author} {\bibfnamefont {K.~D.}\ \bibnamefont
  {J{\"o}ns}}, \bibinfo {author} {\bibfnamefont {M.}~\bibnamefont
  {Gl{\"a}ssl}}, \ and\ \bibinfo {author} {\bibfnamefont {P.}~\bibnamefont
  {Michler}},\ }\href {https://doi.org/10.1038/nphoton.2013.377} {\bibfield
  {journal} {\bibinfo  {journal} {Nat. Photonics}\ }\textbf {\bibinfo {volume}
  {8}},\ \bibinfo {pages} {224} (\bibinfo {year} {2014})}\BibitemShut {NoStop}%
\bibitem [{\citenamefont {Fognini}\ \emph {et~al.}(2019)\citenamefont
  {Fognini}, \citenamefont {Ahmadi}, \citenamefont {Zeeshan}, \citenamefont
  {Fokkens}, \citenamefont {Gibson}, \citenamefont {Sherlekar}, \citenamefont
  {Daley}, \citenamefont {Dalacu}, \citenamefont {Poole}, \citenamefont
  {J\"ons}, \citenamefont {Zwiller},\ and\ \citenamefont
  {Reimer}}]{Fognini_2019}%
  \BibitemOpen
  \bibfield  {author} {\bibinfo {author} {\bibfnamefont {A.}~\bibnamefont
  {Fognini}}, \bibinfo {author} {\bibfnamefont {A.}~\bibnamefont {Ahmadi}},
  \bibinfo {author} {\bibfnamefont {M.}~\bibnamefont {Zeeshan}}, \bibinfo
  {author} {\bibfnamefont {J.~T.}\ \bibnamefont {Fokkens}}, \bibinfo {author}
  {\bibfnamefont {S.~J.}\ \bibnamefont {Gibson}}, \bibinfo {author}
  {\bibfnamefont {N.}~\bibnamefont {Sherlekar}}, \bibinfo {author}
  {\bibfnamefont {S.~J.}\ \bibnamefont {Daley}}, \bibinfo {author}
  {\bibfnamefont {D.}~\bibnamefont {Dalacu}}, \bibinfo {author} {\bibfnamefont
  {P.~J.}\ \bibnamefont {Poole}}, \bibinfo {author} {\bibfnamefont {K.~D.}\
  \bibnamefont {J\"ons}}, \bibinfo {author} {\bibfnamefont {V.}~\bibnamefont
  {Zwiller}}, \ and\ \bibinfo {author} {\bibfnamefont {M.~E.}\ \bibnamefont
  {Reimer}},\ }\href {\doibase 10.1021/acsphotonics.8b01496} {\bibfield
  {journal} {\bibinfo  {journal} {ACS Photonics}\ }\textbf {\bibinfo {volume}
  {6}},\ \bibinfo {pages} {1656} (\bibinfo {year} {2019})}\BibitemShut
  {NoStop}%
\bibitem [{\citenamefont {Akopian}\ \emph {et~al.}(2006)\citenamefont
  {Akopian}, \citenamefont {Lindner}, \citenamefont {Poem}, \citenamefont
  {Berlatzky}, \citenamefont {Avron}, \citenamefont {Gershoni}, \citenamefont
  {Gerardot},\ and\ \citenamefont {Petroff}}]{entangled-photon2}%
  \BibitemOpen
  \bibfield  {author} {\bibinfo {author} {\bibfnamefont {N.}~\bibnamefont
  {Akopian}}, \bibinfo {author} {\bibfnamefont {N.~H.}\ \bibnamefont
  {Lindner}}, \bibinfo {author} {\bibfnamefont {E.}~\bibnamefont {Poem}},
  \bibinfo {author} {\bibfnamefont {Y.}~\bibnamefont {Berlatzky}}, \bibinfo
  {author} {\bibfnamefont {J.}~\bibnamefont {Avron}}, \bibinfo {author}
  {\bibfnamefont {D.}~\bibnamefont {Gershoni}}, \bibinfo {author}
  {\bibfnamefont {B.~D.}\ \bibnamefont {Gerardot}}, \ and\ \bibinfo {author}
  {\bibfnamefont {P.~M.}\ \bibnamefont {Petroff}},\ }\href {\doibase
  10.1103/PhysRevLett.96.130501} {\bibfield  {journal} {\bibinfo  {journal}
  {Phys. Rev. Lett.}\ }\textbf {\bibinfo {volume} {96}},\ \bibinfo {pages}
  {130501} (\bibinfo {year} {2006})}\BibitemShut {NoStop}%
\bibitem [{\citenamefont {Hafenbrak}\ \emph {et~al.}(2007)\citenamefont
  {Hafenbrak}, \citenamefont {Ulrich}, \citenamefont {Michler}, \citenamefont
  {Wang}, \citenamefont {Rastelli},\ and\ \citenamefont {Schmidt}}]{Hafenbrak}%
  \BibitemOpen
  \bibfield  {author} {\bibinfo {author} {\bibfnamefont {R.}~\bibnamefont
  {Hafenbrak}}, \bibinfo {author} {\bibfnamefont {S.~M.}\ \bibnamefont
  {Ulrich}}, \bibinfo {author} {\bibfnamefont {P.}~\bibnamefont {Michler}},
  \bibinfo {author} {\bibfnamefont {L.}~\bibnamefont {Wang}}, \bibinfo {author}
  {\bibfnamefont {A.}~\bibnamefont {Rastelli}}, \ and\ \bibinfo {author}
  {\bibfnamefont {O.~G.}\ \bibnamefont {Schmidt}},\ }\href
  {http://stacks.iop.org/1367-2630/9/i=9/a=315} {\bibfield  {journal} {\bibinfo
   {journal} {New J. Phys.}\ }\textbf {\bibinfo {volume} {9}},\ \bibinfo
  {pages} {315} (\bibinfo {year} {2007})}\BibitemShut {NoStop}%
\bibitem [{\citenamefont {Bennett}\ \emph {et~al.}(2010)\citenamefont
  {Bennett}, \citenamefont {Pooley}, \citenamefont {Stevenson}, \citenamefont
  {Ward}, \citenamefont {Patel}, \citenamefont {{Boyer de la Giroday}},
  \citenamefont {Sk{\"o}ld}, \citenamefont {Farrer}, \citenamefont {Nicoll},
  \citenamefont {Ritchie},\ and\ \citenamefont
  {Shields}}]{Biexc_FSS_electrical_control_Bennett}%
  \BibitemOpen
  \bibfield  {author} {\bibinfo {author} {\bibfnamefont {A.~J.}\ \bibnamefont
  {Bennett}}, \bibinfo {author} {\bibfnamefont {M.~A.}\ \bibnamefont {Pooley}},
  \bibinfo {author} {\bibfnamefont {R.~M.}\ \bibnamefont {Stevenson}}, \bibinfo
  {author} {\bibfnamefont {M.~B.}\ \bibnamefont {Ward}}, \bibinfo {author}
  {\bibfnamefont {R.~B.}\ \bibnamefont {Patel}}, \bibinfo {author}
  {\bibfnamefont {A.}~\bibnamefont {{Boyer de la Giroday}}}, \bibinfo {author}
  {\bibfnamefont {N.}~\bibnamefont {Sk{\"o}ld}}, \bibinfo {author}
  {\bibfnamefont {I.}~\bibnamefont {Farrer}}, \bibinfo {author} {\bibfnamefont
  {C.~A.}\ \bibnamefont {Nicoll}}, \bibinfo {author} {\bibfnamefont {D.~A.}\
  \bibnamefont {Ritchie}}, \ and\ \bibinfo {author} {\bibfnamefont {A.~J.}\
  \bibnamefont {Shields}},\ }\href {\doibase 10.1038/nphys1780} {\bibfield
  {journal} {\bibinfo  {journal} {Nat. Phys.}\ }\textbf {\bibinfo {volume}
  {6}},\ \bibinfo {pages} {947} (\bibinfo {year} {2010})}\BibitemShut {NoStop}%
\bibitem [{\citenamefont {del Valle}(2013)}]{EdV}%
  \BibitemOpen
  \bibfield  {author} {\bibinfo {author} {\bibfnamefont {E.}~\bibnamefont {del
  Valle}},\ }\href {http://stacks.iop.org/1367-2630/15/i=2/a=025019} {\bibfield
   {journal} {\bibinfo  {journal} {New J. Phys.}\ }\textbf {\bibinfo {volume}
  {15}},\ \bibinfo {pages} {025019} (\bibinfo {year} {2013})}\BibitemShut
  {NoStop}%
\bibitem [{\citenamefont {Troiani}, \citenamefont {Perea},\ and\ \citenamefont
  {Tejedor}(2006)}]{Troiani2006}%
  \BibitemOpen
  \bibfield  {author} {\bibinfo {author} {\bibfnamefont {F.}~\bibnamefont
  {Troiani}}, \bibinfo {author} {\bibfnamefont {J.~I.}\ \bibnamefont {Perea}},
  \ and\ \bibinfo {author} {\bibfnamefont {C.}~\bibnamefont {Tejedor}},\ }\href
  {\doibase 10.1103/PhysRevB.74.235310} {\bibfield  {journal} {\bibinfo
  {journal} {Phys. Rev. B}\ }\textbf {\bibinfo {volume} {74}},\ \bibinfo
  {pages} {235310} (\bibinfo {year} {2006})}\BibitemShut {NoStop}%
\bibitem [{\citenamefont {Stevenson}\ \emph {et~al.}(2012)\citenamefont
  {Stevenson}, \citenamefont {Salter}, \citenamefont {Nilsson}, \citenamefont
  {Bennett}, \citenamefont {Ward}, \citenamefont {Farrer}, \citenamefont
  {Ritchie},\ and\ \citenamefont {Shields}}]{stevenson:2012}%
  \BibitemOpen
  \bibfield  {author} {\bibinfo {author} {\bibfnamefont {R.~M.}\ \bibnamefont
  {Stevenson}}, \bibinfo {author} {\bibfnamefont {C.~L.}\ \bibnamefont
  {Salter}}, \bibinfo {author} {\bibfnamefont {J.}~\bibnamefont {Nilsson}},
  \bibinfo {author} {\bibfnamefont {A.~J.}\ \bibnamefont {Bennett}}, \bibinfo
  {author} {\bibfnamefont {M.~B.}\ \bibnamefont {Ward}}, \bibinfo {author}
  {\bibfnamefont {I.}~\bibnamefont {Farrer}}, \bibinfo {author} {\bibfnamefont
  {D.~A.}\ \bibnamefont {Ritchie}}, \ and\ \bibinfo {author} {\bibfnamefont
  {A.~J.}\ \bibnamefont {Shields}},\ }\href {\doibase
  10.1103/PhysRevLett.108.040503} {\bibfield  {journal} {\bibinfo  {journal}
  {Phys. Rev. Lett.}\ }\textbf {\bibinfo {volume} {108}},\ \bibinfo {pages}
  {040503} (\bibinfo {year} {2012})}\BibitemShut {NoStop}%
\bibitem [{\citenamefont {Benson}\ \emph {et~al.}(2000)\citenamefont {Benson},
  \citenamefont {Santori}, \citenamefont {Pelton},\ and\ \citenamefont
  {Yamamoto}}]{Benson_2000_QD_cav_device}%
  \BibitemOpen
  \bibfield  {author} {\bibinfo {author} {\bibfnamefont {O.}~\bibnamefont
  {Benson}}, \bibinfo {author} {\bibfnamefont {C.}~\bibnamefont {Santori}},
  \bibinfo {author} {\bibfnamefont {M.}~\bibnamefont {Pelton}}, \ and\ \bibinfo
  {author} {\bibfnamefont {Y.}~\bibnamefont {Yamamoto}},\ }\href {\doibase
  10.1103/PhysRevLett.84.2513} {\bibfield  {journal} {\bibinfo  {journal}
  {Phys. Rev. Lett.}\ }\textbf {\bibinfo {volume} {84}},\ \bibinfo {pages}
  {2513} (\bibinfo {year} {2000})}\BibitemShut {NoStop}%
\bibitem [{\citenamefont {S{\'a}nchez~Mu{\~n}oz}\ \emph
  {et~al.}(2015)\citenamefont {S{\'a}nchez~Mu{\~n}oz}, \citenamefont {Laussy},
  \citenamefont {Tejedor},\ and\ \citenamefont {del Valle}}]{munoz15}%
  \BibitemOpen
  \bibfield  {author} {\bibinfo {author} {\bibfnamefont {C.}~\bibnamefont
  {S{\'a}nchez~Mu{\~n}oz}}, \bibinfo {author} {\bibfnamefont {F.~P.}\
  \bibnamefont {Laussy}}, \bibinfo {author} {\bibfnamefont {C.}~\bibnamefont
  {Tejedor}}, \ and\ \bibinfo {author} {\bibfnamefont {E.}~\bibnamefont {del
  Valle}},\ }\href {http://stacks.iop.org/1367-2630/17/i=12/a=123021}
  {\bibfield  {journal} {\bibinfo  {journal} {New J. Phys.}\ }\textbf {\bibinfo
  {volume} {17}},\ \bibinfo {pages} {123021} (\bibinfo {year}
  {2015})}\BibitemShut {NoStop}%
\bibitem [{\citenamefont {Seidelmann}\ \emph {et~al.}(2021)\citenamefont
  {Seidelmann}, \citenamefont {Cosacchi}, \citenamefont {Cygorek},
  \citenamefont {Reiter}, \citenamefont {Vagov},\ and\ \citenamefont
  {Axt}}]{Seidelmann_QUTE_2020}%
  \BibitemOpen
  \bibfield  {author} {\bibinfo {author} {\bibfnamefont {T.}~\bibnamefont
  {Seidelmann}}, \bibinfo {author} {\bibfnamefont {M.}~\bibnamefont
  {Cosacchi}}, \bibinfo {author} {\bibfnamefont {M.}~\bibnamefont {Cygorek}},
  \bibinfo {author} {\bibfnamefont {D.~E.}\ \bibnamefont {Reiter}}, \bibinfo
  {author} {\bibfnamefont {A.}~\bibnamefont {Vagov}}, \ and\ \bibinfo {author}
  {\bibfnamefont {V.~M.}\ \bibnamefont {Axt}},\ }\href {\doibase
  https://doi.org/10.1002/qute.202000108} {\bibfield  {journal} {\bibinfo
  {journal} {Adv. Quantum Technol.}\ }\textbf {\bibinfo {volume} {4}},\
  \bibinfo {pages} {2000108} (\bibinfo {year} {2021})}\BibitemShut {NoStop}%
\bibitem [{\citenamefont {Ardelt}\ \emph {et~al.}(2016)\citenamefont {Ardelt},
  \citenamefont {Koller}, \citenamefont {Simmet}, \citenamefont {Hanschke},
  \citenamefont {Bechtold}, \citenamefont {Regler}, \citenamefont
  {Wierzbowski}, \citenamefont {Riedl}, \citenamefont {Finley},\ and\
  \citenamefont {M\"uller}}]{Ardelt_exp_const_driving}%
  \BibitemOpen
  \bibfield  {author} {\bibinfo {author} {\bibfnamefont {P.-L.}\ \bibnamefont
  {Ardelt}}, \bibinfo {author} {\bibfnamefont {M.}~\bibnamefont {Koller}},
  \bibinfo {author} {\bibfnamefont {T.}~\bibnamefont {Simmet}}, \bibinfo
  {author} {\bibfnamefont {L.}~\bibnamefont {Hanschke}}, \bibinfo {author}
  {\bibfnamefont {A.}~\bibnamefont {Bechtold}}, \bibinfo {author}
  {\bibfnamefont {A.}~\bibnamefont {Regler}}, \bibinfo {author} {\bibfnamefont
  {J.}~\bibnamefont {Wierzbowski}}, \bibinfo {author} {\bibfnamefont
  {H.}~\bibnamefont {Riedl}}, \bibinfo {author} {\bibfnamefont {J.~J.}\
  \bibnamefont {Finley}}, \ and\ \bibinfo {author} {\bibfnamefont
  {K.}~\bibnamefont {M\"uller}},\ }\href {\doibase 10.1103/PhysRevB.93.165305}
  {\bibfield  {journal} {\bibinfo  {journal} {Phys. Rev. B}\ }\textbf {\bibinfo
  {volume} {93}},\ \bibinfo {pages} {165305} (\bibinfo {year}
  {2016})}\BibitemShut {NoStop}%
\bibitem [{\citenamefont {Hargart}\ \emph {et~al.}(2016)\citenamefont
  {Hargart}, \citenamefont {M\"uller}, \citenamefont {Roy-Choudhury},
  \citenamefont {Portalupi}, \citenamefont {Schneider}, \citenamefont
  {H\"ofling}, \citenamefont {Kamp}, \citenamefont {Hughes},\ and\
  \citenamefont {Michler}}]{Hargart_exp_const_driving}%
  \BibitemOpen
  \bibfield  {author} {\bibinfo {author} {\bibfnamefont {F.}~\bibnamefont
  {Hargart}}, \bibinfo {author} {\bibfnamefont {M.}~\bibnamefont {M\"uller}},
  \bibinfo {author} {\bibfnamefont {K.}~\bibnamefont {Roy-Choudhury}}, \bibinfo
  {author} {\bibfnamefont {S.~L.}\ \bibnamefont {Portalupi}}, \bibinfo {author}
  {\bibfnamefont {C.}~\bibnamefont {Schneider}}, \bibinfo {author}
  {\bibfnamefont {S.}~\bibnamefont {H\"ofling}}, \bibinfo {author}
  {\bibfnamefont {M.}~\bibnamefont {Kamp}}, \bibinfo {author} {\bibfnamefont
  {S.}~\bibnamefont {Hughes}}, \ and\ \bibinfo {author} {\bibfnamefont
  {P.}~\bibnamefont {Michler}},\ }\href {\doibase 10.1103/PhysRevB.93.115308}
  {\bibfield  {journal} {\bibinfo  {journal} {Phys. Rev. B}\ }\textbf {\bibinfo
  {volume} {93}},\ \bibinfo {pages} {115308} (\bibinfo {year}
  {2016})}\BibitemShut {NoStop}%
\bibitem [{\citenamefont {Mermillod}\ \emph {et~al.}(2016)\citenamefont
  {Mermillod}, \citenamefont {Wigger}, \citenamefont {Delmonte}, \citenamefont
  {Reiter}, \citenamefont {Schneider}, \citenamefont {Kamp}, \citenamefont
  {H\"{o}fling}, \citenamefont {Langbein}, \citenamefont {Kuhn}, \citenamefont
  {Nogues},\ and\ \citenamefont {Kasprzak}}]{Mermillod:16}%
  \BibitemOpen
  \bibfield  {author} {\bibinfo {author} {\bibfnamefont {Q.}~\bibnamefont
  {Mermillod}}, \bibinfo {author} {\bibfnamefont {D.}~\bibnamefont {Wigger}},
  \bibinfo {author} {\bibfnamefont {V.}~\bibnamefont {Delmonte}}, \bibinfo
  {author} {\bibfnamefont {D.~E.}\ \bibnamefont {Reiter}}, \bibinfo {author}
  {\bibfnamefont {C.}~\bibnamefont {Schneider}}, \bibinfo {author}
  {\bibfnamefont {M.}~\bibnamefont {Kamp}}, \bibinfo {author} {\bibfnamefont
  {S.}~\bibnamefont {H\"{o}fling}}, \bibinfo {author} {\bibfnamefont
  {W.}~\bibnamefont {Langbein}}, \bibinfo {author} {\bibfnamefont
  {T.}~\bibnamefont {Kuhn}}, \bibinfo {author} {\bibfnamefont {G.}~\bibnamefont
  {Nogues}}, \ and\ \bibinfo {author} {\bibfnamefont {J.}~\bibnamefont
  {Kasprzak}},\ }\href {\doibase 10.1364/OPTICA.3.000377} {\bibfield  {journal}
  {\bibinfo  {journal} {Optica}\ }\textbf {\bibinfo {volume} {3}},\ \bibinfo
  {pages} {377} (\bibinfo {year} {2016})}\BibitemShut {NoStop}%
\bibitem [{\citenamefont {Reiter}\ \emph {et~al.}(2014)\citenamefont {Reiter},
  \citenamefont {Kuhn}, \citenamefont {Gl\"assl},\ and\ \citenamefont
  {Axt}}]{Reiter_2014}%
  \BibitemOpen
  \bibfield  {author} {\bibinfo {author} {\bibfnamefont {D.~E.}\ \bibnamefont
  {Reiter}}, \bibinfo {author} {\bibfnamefont {T.}~\bibnamefont {Kuhn}},
  \bibinfo {author} {\bibfnamefont {M.}~\bibnamefont {Gl\"assl}}, \ and\
  \bibinfo {author} {\bibfnamefont {V.~M.}\ \bibnamefont {Axt}},\ }\href
  {\doibase 10.1088/0953-8984/26/42/423203} {\bibfield  {journal} {\bibinfo
  {journal} {J. Phys.: Condens. Matter}\ }\textbf {\bibinfo {volume} {26}},\
  \bibinfo {pages} {423203} (\bibinfo {year} {2014})}\BibitemShut {NoStop}%
\bibitem [{\citenamefont {Reindl}\ \emph {et~al.}(2017)\citenamefont {Reindl},
  \citenamefont {J\"ons}, \citenamefont {Huber}, \citenamefont {Schimpf},
  \citenamefont {Huo}, \citenamefont {Zwiller}, \citenamefont {Rastelli},\ and\
  \citenamefont {Trotta}}]{reindl2017}%
  \BibitemOpen
  \bibfield  {author} {\bibinfo {author} {\bibfnamefont {M.}~\bibnamefont
  {Reindl}}, \bibinfo {author} {\bibfnamefont {K.~D.}\ \bibnamefont {J\"ons}},
  \bibinfo {author} {\bibfnamefont {D.}~\bibnamefont {Huber}}, \bibinfo
  {author} {\bibfnamefont {C.}~\bibnamefont {Schimpf}}, \bibinfo {author}
  {\bibfnamefont {Y.}~\bibnamefont {Huo}}, \bibinfo {author} {\bibfnamefont
  {V.}~\bibnamefont {Zwiller}}, \bibinfo {author} {\bibfnamefont
  {A.}~\bibnamefont {Rastelli}}, \ and\ \bibinfo {author} {\bibfnamefont
  {R.}~\bibnamefont {Trotta}},\ }\href {\doibase 10.1021/acs.nanolett.7b00777}
  {\bibfield  {journal} {\bibinfo  {journal} {Nano Lett.}\ }\textbf {\bibinfo
  {volume} {17}},\ \bibinfo {pages} {4090} (\bibinfo {year}
  {2017})}\BibitemShut {NoStop}%
\bibitem [{\citenamefont {Hanschke}\ \emph {et~al.}(2018)\citenamefont
  {Hanschke}, \citenamefont {Fischer}, \citenamefont {Appel}, \citenamefont
  {Lukin}, \citenamefont {Wierzbowski}, \citenamefont {Sun}, \citenamefont
  {Trivedi}, \citenamefont {Vuckovi{\'c}}, \citenamefont {Finley},\ and\
  \citenamefont {M{\"u}ller}}]{hanschke2018}%
  \BibitemOpen
  \bibfield  {author} {\bibinfo {author} {\bibfnamefont {L.}~\bibnamefont
  {Hanschke}}, \bibinfo {author} {\bibfnamefont {K.~A.}\ \bibnamefont
  {Fischer}}, \bibinfo {author} {\bibfnamefont {S.}~\bibnamefont {Appel}},
  \bibinfo {author} {\bibfnamefont {D.}~\bibnamefont {Lukin}}, \bibinfo
  {author} {\bibfnamefont {J.}~\bibnamefont {Wierzbowski}}, \bibinfo {author}
  {\bibfnamefont {S.}~\bibnamefont {Sun}}, \bibinfo {author} {\bibfnamefont
  {R.}~\bibnamefont {Trivedi}}, \bibinfo {author} {\bibfnamefont
  {J.}~\bibnamefont {Vuckovi{\'c}}}, \bibinfo {author} {\bibfnamefont {J.~J.}\
  \bibnamefont {Finley}}, \ and\ \bibinfo {author} {\bibfnamefont
  {K.}~\bibnamefont {M{\"u}ller}},\ }\href {\doibase 10.1038/s41534-018-0092-0}
  {\bibfield  {journal} {\bibinfo  {journal} {npj Quantum Inf.}\ }\textbf
  {\bibinfo {volume} {4}},\ \bibinfo {pages} {43} (\bibinfo {year}
  {2018})}\BibitemShut {NoStop}%
\bibitem [{\citenamefont {Lindblad}(1976)}]{Lindblad:1976}%
  \BibitemOpen
  \bibfield  {author} {\bibinfo {author} {\bibfnamefont {G.}~\bibnamefont
  {Lindblad}},\ }\href {\doibase 10.1007/BF01608499} {\bibfield  {journal}
  {\bibinfo  {journal} {Commun. Math. Phys.}\ }\textbf {\bibinfo {volume}
  {48}},\ \bibinfo {pages} {119} (\bibinfo {year} {1976})}\BibitemShut
  {NoStop}%
\bibitem [{\citenamefont {Cosacchi}\ \emph {et~al.}(2018)\citenamefont
  {Cosacchi}, \citenamefont {Cygorek}, \citenamefont {Ungar}, \citenamefont
  {Barth}, \citenamefont {Vagov},\ and\ \citenamefont {Axt}}]{multi-time}%
  \BibitemOpen
  \bibfield  {author} {\bibinfo {author} {\bibfnamefont {M.}~\bibnamefont
  {Cosacchi}}, \bibinfo {author} {\bibfnamefont {M.}~\bibnamefont {Cygorek}},
  \bibinfo {author} {\bibfnamefont {F.}~\bibnamefont {Ungar}}, \bibinfo
  {author} {\bibfnamefont {A.~M.}\ \bibnamefont {Barth}}, \bibinfo {author}
  {\bibfnamefont {A.}~\bibnamefont {Vagov}}, \ and\ \bibinfo {author}
  {\bibfnamefont {V.~M.}\ \bibnamefont {Axt}},\ }\href
  {https://link.aps.org/doi/10.1103/PhysRevB.98.125302} {\bibfield  {journal}
  {\bibinfo  {journal} {Phys. Rev. B}\ }\textbf {\bibinfo {volume} {98}},\
  \bibinfo {pages} {125302} (\bibinfo {year} {2018})}\BibitemShut {NoStop}%
\bibitem [{Note1()}]{Note1}%
  \BibitemOpen
  \bibinfo {note} {Note that in typical experiments the measurements are
  performed on photons which have already left the cavity. Nevertheless, when
  the out-coupling of light out of the cavity is considered to be a Markovian
  process, Eq.~\protect \textup {\hbox {\mathsurround \z@ \protect \normalfont
  (\ignorespaces \ref {eq:G2}\unskip \@@italiccorr )}} can be used to describe
  $G^{(2)}(t,\tau )$ as measured outside of the cavity [cf. Refs.~\protect
  \rev@citealpnum {Kuhn:16,Different-Concurrences:18}].}\BibitemShut {Stop}%
\bibitem [{\citenamefont {James}\ \emph {et~al.}(2001)\citenamefont {James},
  \citenamefont {Kwiat}, \citenamefont {Munro},\ and\ \citenamefont
  {White}}]{QuantumStateTomography}%
  \BibitemOpen
  \bibfield  {author} {\bibinfo {author} {\bibfnamefont {D.~F.~V.}\
  \bibnamefont {James}}, \bibinfo {author} {\bibfnamefont {P.~G.}\ \bibnamefont
  {Kwiat}}, \bibinfo {author} {\bibfnamefont {W.~J.}\ \bibnamefont {Munro}}, \
  and\ \bibinfo {author} {\bibfnamefont {A.~G.}\ \bibnamefont {White}},\ }\href
  {\doibase 10.1103/PhysRevA.64.052312} {\bibfield  {journal} {\bibinfo
  {journal} {Phys. Rev. A}\ }\textbf {\bibinfo {volume} {64}},\ \bibinfo
  {pages} {052312} (\bibinfo {year} {2001})}\BibitemShut {NoStop}%
\bibitem [{\citenamefont {Wootters}(1998)}]{Wootters1998}%
  \BibitemOpen
  \bibfield  {author} {\bibinfo {author} {\bibfnamefont {W.~K.}\ \bibnamefont
  {Wootters}},\ }\href {\doibase 10.1103/PhysRevLett.80.2245} {\bibfield
  {journal} {\bibinfo  {journal} {Phys. Rev. Lett.}\ }\textbf {\bibinfo
  {volume} {80}},\ \bibinfo {pages} {2245} (\bibinfo {year}
  {1998})}\BibitemShut {NoStop}%
\bibitem [{Note2()}]{Note2}%
  \BibitemOpen
  \bibinfo {note} { $C=\max\left\lbrace \sqrt{\lambda_1}-\sqrt{\lambda_2}-\sqrt{\lambda_3}-\sqrt{\lambda_4},0\right\rbrace$ where $\lambda_j\geq\lambda_{j+1}$ are the eigenvalues of $\rho^\text{2p}T{\rho^\text{2p}}^\ast T$ in decreasing order and $T$ is the antidiagonal matrix with elements $\left\lbrace -1,1,1,-1\right\rbrace$.}
  \BibitemShut {NoStop}%
\bibitem [{\citenamefont {Stevenson}\ \emph {et~al.}(2008)\citenamefont
  {Stevenson}, \citenamefont {Hudson}, \citenamefont {Bennett}, \citenamefont
  {Young}, \citenamefont {Nicoll}, \citenamefont {Ritchie},\ and\ \citenamefont
  {Shields}}]{StevensonPRL2008}%
  \BibitemOpen
  \bibfield  {author} {\bibinfo {author} {\bibfnamefont {R.~M.}\ \bibnamefont
  {Stevenson}}, \bibinfo {author} {\bibfnamefont {A.~J.}\ \bibnamefont
  {Hudson}}, \bibinfo {author} {\bibfnamefont {A.~J.}\ \bibnamefont {Bennett}},
  \bibinfo {author} {\bibfnamefont {R.~J.}\ \bibnamefont {Young}}, \bibinfo
  {author} {\bibfnamefont {C.~A.}\ \bibnamefont {Nicoll}}, \bibinfo {author}
  {\bibfnamefont {D.~A.}\ \bibnamefont {Ritchie}}, \ and\ \bibinfo {author}
  {\bibfnamefont {A.~J.}\ \bibnamefont {Shields}},\ }\href {\doibase
  10.1103/PhysRevLett.101.170501} {\bibfield  {journal} {\bibinfo  {journal}
  {Phys. Rev. Lett.}\ }\textbf {\bibinfo {volume} {101}},\ \bibinfo {pages}
  {170501} (\bibinfo {year} {2008})}\BibitemShut {NoStop}%
\bibitem [{Note3()}]{Note3}%
  \BibitemOpen
  \bibinfo {note} {Note that in the steady state situation with constant
  continuous driving, $\rho ^\protect \text {2p}$ and $C$ do not depend on
  $\Delta t$ since the statistical operator and the Hamiltonian $\protect
  \mathaccentV {hat}05E{H}$ are constant during the measurement
  process.}\BibitemShut {Stop}%
\end{thebibliography}

%

\end{document}